\documentclass[graybox]{svmult}
\usepackage{mathptmx}       
\usepackage{helvet}         
\usepackage{courier}        
\usepackage{makeidx}         
\usepackage{graphicx}        
\usepackage{multicol}        
\usepackage[bottom]{footmisc}
\usepackage{amsmath,amssymb,amsfonts}        
\makeindex             

\usepackage{bm}
\usepackage{bbm}
\usepackage{verbatim}

\def\alert#1{\textcolor{black}{#1}}

\def\alertB#1{\textcolor{black}{#1}}

\begin{document}
\title*{Electrostatic Computations for Statistical Mechanics and Random Matrix Applications}
\author{Sung-Soo Byun and Peter J. Forrester}
\institute{Sung-Soo Byun \at Department of Mathematical Sciences and Research Institute of Mathematics, Seoul National University, Seoul 151-747, Republic of Korea, \email{sungsoobyun@snu.ac.kr}
\and Peter J. Forrester \at School of Mathematics and Statistics,  The University of Melbourne,
Victoria 3010, Australia, \email{pjforr@unimelb.edu.au}}
%
%
\maketitle 

\abstract{Although for the most part classical, the topic of electrostatics finds to this day new applications. In this review we highlight several theoretical results on electrostatics, chosen to both illustrate general principles, and for their application in statistical mechanics and random matrix settings. The theoretical results include electrostatic potentials and energies associated with balls and hyperellipsoids in general dimension, equilibrium measures associated with surfaces of the latter, the use of conformal mappings in two-dimensions, and the balayage measure. A number of explicit examples of their use in predicting the leading asymptotic form of certain configuration integrals and particle density in particular statistical mechanical systems are given, as well as with regards to questions relating to fluctuation formulas and (conditioned) gap probabilities.
}


\section{Scope of the Review}
\subsection{Coulomb Potentials and the One-Component Plasma Model}\label{S1.1}
Consider two unit charges in $\mathbb R^d$ at positions $\vec{r},\vec{r}\,'$, and denote by $\Phi_d(\vec{r},\vec{r}\,')$ the corresponding potential. We will take the latter as being specified by the solution of the particular $d$-dimensional, rescaled Poisson equation
\begin{equation}\label{1.1}
\nabla_{\vec{r}}^2 \Phi_d
(\vec{r},\vec{r}\,') =
- c_d \chi_d \delta(\vec{r} - \vec{r}\,'),
\end{equation}
where $c_d:= 2 \pi^{d/2}/\Gamma(d/2)$ is the surface area of the unit sphere $|\vec{r}|^2 = 1$ embedded in $\mathbb R^d$, and
$\chi_d = d-2$ for $d > 2$ and equals unity otherwise.
Satisfying (\ref{1.1}) are the so-called free boundary conditions solutions
\begin{equation}\label{1.2}
\Phi_d
(\vec{r},\vec{r}\,') =
\begin{cases} -  |x - x'|, & d = 1, \\
- \log |\vec{r}- \vec{r}\,'|, & d = 2, \\
|\vec{r}- \vec{r}\,'|^{2 - d},
& d > 2
\end{cases}
\end{equation}
(in the $d=1$ case we have rewritten $\vec{r},\vec{r}\,'$ as $x,x'$ since the positions are on the real line).

Let $C(\vec{r}\,')$ denote a charge density with support $\Omega \subset \mathbb R^d$ (it is convenient to consider $\Omega$ as being open and its closure
$\bar{\Omega}$ as being compact with a smooth boundary). The electrostatic potential at
$\vec{r}$, $V(\vec{r})$ say, of this charge density is given in terms of $\Phi_d
(\vec{r},\vec{r}\,')$ by
\begin{equation}\label{1.3}
V(\vec{r}) = 
\int_\Omega \Phi_d
(\vec{r},\vec{r}\,') C(\vec{r}\,')
\, d \vec{r}\,'.
\end{equation}
Applying $\nabla_{\vec{r}}^2$ to both sides of (\ref{1.3}), commuting this with the integral on the right hand side (RHS) shows
$\nabla_{\vec{r}}^2 V(\vec{r})$ is equal to the RHS of (\ref{1.1}) with $\delta(\vec{r} -\vec{r}\,')$
replaced by $C(\vec{r})$ in the case that $\vec{r} \in \Omega$, and equal to zero for $\vec{r} \in \mathbb R^d \backslash \bar{\Omega}$. This is another example of a Poisson equation. Suppose we now specialise to the choice that for
$\vec{r} \in \Omega$, $C(\vec{r}) = - \rho_{\rm b}$ a constant. Then up to distinct additive harmonic functions
in the two cases, the solution of this new Poisson equation is
\begin{equation}\label{1.4}
V(\vec{r}) = 
\begin{cases}\displaystyle
\rho_{\rm b} \Big ({c_d \chi_d \over 2 d} \Big ) |\vec{r}|^2, &
\vec{r} \in \Omega, \\[.2cm]
0, & {\rm otherwise}
\end{cases}
\end{equation}
(the additive harmonic functions must be such that $V(\vec{r})$ is continuous on $\partial \Omega$, the boundary of $\Omega$). 

The potential $V(\vec{r})$ is a key quantity in the specification of the one-component plasma model of the equilibrium statistical mechanics of a classical Coulomb system \cite{BH80}. For a given domain $\Omega$ in dimension $d$, this model consists of a smeared out, uniform, negatively charged background (charge density $-\rho_{\rm b}$ say) such that
\begin{equation}\label{1.5}
\rho_{\rm b} |\Omega| = N
\end{equation}
(here $|\Omega|$ denotes the volume of $\Omega$). Interacting with the background, and among themselves, are $N$ mobile particles of unit positive charge.
One sees immediately that the potential energy experienced by a unit positive charge at point $\vec r$ is given by the first case of (\ref{1.4}), with the understanding that in the latter the additive harmonic function is still to be determined (and will depend on $\Omega$). Thus with the potential energy $U_{\text{p-b}}$ due to the interaction of the $N$ unit positive charges with the background we have
\begin{equation}\label{1.6}
U_{\text{p-b}} = \sum_{j=1}^N
V(\vec{r}_j).
\end{equation}
This potential is also relevant to the computation of the potential energy  $U_{\text{b-b}}$ of the self interaction of the background. Thus by \eqref{1.3} with $C(\vec{r}) = - \rho_{\rm b}$,
\begin{equation}\label{1.6a}
U_{\text{b-b}} = 
{1 \over 2} \rho_{\rm b}^2 
\int_\Omega d \vec{r}\,
\int_\Omega d \vec{r} \, ' \,
\Phi_d
(\vec{r},\vec{r}\,') =
- {1 \over 2} \rho_{\rm b} 
\int_\Omega V(\vec{r}) \, d \vec{r}
\end{equation}
(here the factor of one half is due to the double counting of the potential energy implied by the double integral).
Noting too that the total potential energy of the interaction between the $N$ unit mobile particles at positions
$\vec{r}_1,\dots, \vec{r}_N$ is 
\begin{equation}
U_{\text{p-p}} := \sum_{1 \le j < k \le N}
\Phi_{\rm d}(\vec{r}_j,\vec{r}_k), 
\end{equation}
one has that the Boltzmann factor for the system is $e^{-\beta (U_{\text{p-p}} + U_{\text{p-b}}+U_{\text{b-b}})}$; see also \cite[\S 1.4.1]{Fo10}, \cite[\S III.A]{Le22}. Here $\beta$ is the dimensionless inverse temperature.

In the formalism of equilibrium statistical mechanics one proceeds from knowledge of the Boltzmann factor to form the partition function
\begin{equation}\label{1.7}
Z_{N,\beta} = {1 \over N!} 
\int_\Omega d \vec{r}_1\cdots
\int_\Omega d \vec{r}_N \,
e^{-\beta (U_{\text{p-p}} + U_{\text{p-b}}+
U_{\text{b-b}})}.
\end{equation}
From the partition function one computes the dimensionless free energy as $\beta F = - \log Z_{N,\beta}$. A significant property of this quantity is that 
for $\Omega= N^{1/d} \Omega_0$ with $|\Omega_0 | = \rho_{\rm b}$ (i.e.~$\Omega$ relates to $\Omega_0$ by a dilation consistent with
(\ref{1.5})), the 
large $N$ limit of
$\beta F/N$ exists. Thus the dimensionless free energy is extensive, being proportional to $N$, and so has a well defined thermodynamic limit \cite{LN75},
\cite{SM76}, \cite[Th.23]{Le22}.

A useful viewpoint in understanding the role played by all terms in the total potential energy is the mean field heuristic
that for typical configurations the density of the particles will \alert{be} equal to that of the background. \alert{C}onsequently, to leading order
for typical configurations,
$$
U_{\text{p-b}} \sim
\rho_{\rm b} \int_\Omega V(\vec{r}) \,
\, d \vec{r},\qquad U_{\text{p-p}} \sim U_{\text{b-b}}.
$$
Accepting this, and recalling (\ref{1.6a}), one is lead to the conclusion that the leading order of the three terms making up the total energy cancels. An alternative viewpoint is that if (\ref{1.7}) was to be considered without $U_{\text{b-b}}$  present in the total energy, one could predict the leading large $N$ form as being the leading large $N$ form of $e^{\beta U_{\text{b-b}} }$.  \alert{Hence the leading asymptotics is}  entirely determined by electrostatics.

This latter conclusion is far from isolated. Asking the question as to what the leading order particle density of the one-component plasma must be, macroscopic electrostatics gives the answer as equal and opposite to the background charge density. If not an electric field would be created and the system would be out of equilibrium. In other directions, in the work \cite{JLM93}, macroscopic electrostatics \alert{is} used to predict the leading form of the tails of the probability distribution for the random variable $N(R)$ counting the number of particles in a ball of radius $R$ for the two and three dimensional one-component plasma. Here it is assumed that the thermodynamic limit has already been taken, and that $R \gg 1$. Note that the mean of this distribution is $\rho_{\rm b} |D_R|$, where $|D_R|$ is the volume of the ball of radius $R$, and the standard deviation is proportional to $|\partial D_R|^{1/2}$ (i.e.~the square root of that surface area of $D_R$) \cite{MY80}.
\alertB{This latter property characterises  an effect referred to as hyperuniformity \cite{To18}.}
The \alert{particular} tails,
or equivalently \alert{particular} large deviation regime,
considered in \cite{JLM93} \alert{is}  the probability that the number of particles present in $D_R$ is greater than $c|\partial D_R|^{\alpha}$ for $\alpha > 1/2$, $c>0$.
\alert{T}he details of this formula in the case $\alpha > 2$
\alert{are presented in} (\ref{5.2d}) below.
Related electrostatic based hole probability calculations in the two-dimensional case, when there is an application to the complex Ginibre random matrix ensemble (see \S \ref{S3.4.2} below), can be found in \cite{Fo93x}, \cite{ATW14}, \cite{AR17}, \cite{Ch23}. \alert{With much recent research involving two-dimensional Coulomb gas models (or closely related random matrix models --- see
subsection \ref{S3.4.2} below)
centred on aspects of the random variable
$N(R)$ (see for example \cite{Ch22, FL22, ABE23, ABES23, ADM24, MMO25, Noda25}), one expects that the work to date along this line where electrostatic based computations are relevant is far from exhausted.}

\subsection{One-Component Riesz Potential Models}\label{S1.2}

Let $X$ be an $N \times N$ random matrix with independent standard real (complex) Gaussian entries, and form from this the Hermitian matrix $H = {1 \over 2} (X + X^\dagger)$. A classical result in random matrix theory is that the joint eigenvalue probability density function (PDF) is proportional to
\begin{equation}\label{1.8}
\prod_{l=1}^N e^{-\beta \lambda_l^2/2} \prod_{1 \le j < k \le N} | \lambda_k - \lambda_j|^\beta,
\quad \lambda_l \in \mathbb R \:
(l=1,\dots,N),
\end{equation}
where $\beta=1$ ($\beta = 2$) in the real (complex) case; see e.g.~\cite[\S 1.2\&1.3]{Fo10}.
One sees that this has the functional form $e^{-\beta U}$ for
\begin{equation}\label{1.9}
U = {1 \over 2} \sum_{l=1}^N \lambda_l^2 -
\sum_{1 \le j < k \le N} \log | \lambda_k - \lambda_j|,
\end{equation}
and thus has the interpretation as the Boltzmann factor for a one-component log-potential Coulomb systems with background-particle potential equal to $\lambda^2/2$
(up to an additive constant). 
\alert{We recall from (\ref{1.2}) that $-\log |\vec{r} - \vec{r} \, '|$ is the Coulomb potential in dimension $d=2$.}
\alert{However} in (\ref{1.8}) the domain is now one-dimensional rather than two-dimensional, which results in different behaviours. As an example, we know \alert{from (\ref{1.4}) that 
for the one-component plasma with $d=2$
a constant background density is responsible for a quadratic background-particle potential.} \alert{In contrast}, a result which can be traced back to Wigner \cite{Wi57a} gives that a background with profile proportional to $(1 - y^2)^{1/2} \mathbbm 1_{|y|<1}$,
$y = \lambda/\sqrt{2N}$ and
which is now referred to as the Wigner semi-circle, implies
the quadratic potential in (\ref{1.9}). Most importantly from an electrostatics point of view is that this background can be determined by solving the potential theoretic Frostman-type
problem
\begin{equation}\label{1.10}
{1 \over 2} \lambda^2 -
\int_{-c}^c \mu(y) \log|\lambda - y| \, dy  \quad \begin{cases} = C, & |y| \le c, \\
\ge C, & |y|>c, \end{cases}
\end{equation}
for the density $\mu(y)$ and its support $(-c,c)$ ($C$ is sometimes referred to as Robin's constant); see \cite{ST97}.

\alert{Generalising (\ref{1.2})}, this suggests introducing the family of potentials, known as Riesz potentials, 
\begin{equation}\label{1.11}
\Psi_s
(\vec{r},\vec{r}\,') =
\begin{cases} -  |\vec{r}- \vec{r}\,' |^{-s}, & s < 0, \\
- \log |\vec{r}- \vec{r}\,'|, &  s = 0, \\
|\vec{r}- \vec{r}\,'|^{-s},
& s > 0,
\end{cases}
\end{equation}
defined in the first instance independent of the dimension. Further consideration does relate the exponent $s$ to the dimension $d$. Thus, to be in a setting that a smeared out background is necessary for thermodynamic stability, we restrict attention to the cases that $\Psi_s$ is not integrable with respect to $\vec{r}\,'$ at infinity. \alert{Then}, as emphasised in \cite[\S I]{Le22}, for typical configurations
with a constant density
$$
\sum_{j=1}^N
\Psi_s(\vec{r}_j,\vec{r}\,')
$$
will not converge as $N \to \infty$. \alert{Rather}, convergence can only be achieved by the introduction of a smeared out background (recall the paragraph below (\ref{1.7})). This happens for $s \le d$ (note that the Coulomb case (\ref{1.2}) corresponds to $s=d-2$). \alert{As an aside},
we remark that for $d=1$ the short-range regime $s>1$ has  been \alert{the subject of}~\cite{ADKKMMS19,HLSS18}, and \alert{we note} that the particular the case $s=2$ exhibits connections with the Calogero--Moser model \cite{AKD19}.

There is still a further restriction to be placed on $s$. A general principle for statistical mechanical systems with a long range pair potential to be thermodynamically stable is that  of screening (see e.g.~\cite[\S 14.1]{Fo10}). Use of linear response theory gives that for this to \alert{hold} the reciprocal of the Fourier transform of the pair potential (interpreted as a generalised function \cite{Li58}) must go to zero and be positive.  This Fourier transform equals \cite[Eq.~(5)]{Le22}
$2^{-s+d/2} (\Gamma((d-s)/2)/
\Gamma(s/2)) |k|^{s -d}$. Its reciprocal going to zero as $|k| \to 0$ then requires $s < d$ (and hence excludes the boundary case \alert{$s=d$ permitted by the considerations of} the above paragraph) and $s$ not equal to a negative even integer, while positivity requires $s > -2.$ 

\subsection{Outline of the Remainder of the Article}

The first topic we consider, which forms the content of \S \ref{S2}, is the role of the background in the class of \alert{permitted} one-component Riesz gases. Here attention is focused on \alert{the background} being required for existence of the thermodynamic limit, and also on its role as an analytic continuation from the integrable case. 
Section \ref{S3} begins by reviewing the settings of balls and hyperellipsoids in $d$-dimensions, filled with a uniform charge, when the electrostatic potential and energy can be computed explicitly. 
Knowledge of explicit electrostatic energies for uniformly charged domains of \alert{the $d=2$ disk and ellipse, and the $d=1$ interval}  is applied in the final subsection of \S \ref{S3}
to give predictions for the leading asymptotics of certain configurational integrals which occur in statistical mechanics and random matrix settings. 
In \S \ref{S4} these geometries are further considered, now from the viewpoint of their surfaces corresponding to an equipotential supporting a surface charge.

As is well known, going back to Wigner \cite{Wi57a} and Dyson \cite{Dy62}, Hermitian and unitary random matrix ensembles give rise to electrostatic problems involving the logarithmic potential with the charge distribution restricted to one-dimension; recall the discussion around (\ref{1.10}). This is similarly true for certain ensembles of non-Hermitian random matrices with the charges occupying a two-dimensional region; this was already noted in the pioneering work on this topic of Ginibre \cite{Gi65}, and is developed systematically in \cite{BF25}, \cite{Fo16}. Fortunately the electrostatic theory associated with the logarithmic potential is more extensive than that in higher dimensions, due to the applicability of complex variable methods. These are first introduced in \S \ref{S3.4A} in the context of deducing the domain potential, then further developed in \S \ref{S4}, where in particular the computation of certain Green functions is reviewed. The final subsection of \S \ref{S4} gives application of this theory to fluctuation formulas and surface correlations.
The final section of the review, \S \ref{S5}, addresses the balayage measure and the hole probability.

\section{Background as an Analytic Continuation}\label{S2}
From the discussion of \S \ref{S1.2}, the rationale for introducing the uniform background is so that the total potential energy will, for typical configurations, be of order $N$ (i.e.~extensive). A related viewpoint is that the potential at a point in the interior of the system will be of order unity.
For configurations forming lattices, and for particular geometries, there are circumstances when these expected consequences can be verified explicitly.  Further, it can be shown that this potential is an analytic continuation in $s$ of the same quantity when no background is necessary.

\subsection{Riesz Gas on a Circle}
The one-dimensional domain of a circle of radius $R$ lends itself to the simplest analysis.
Due to the rotational symmetry, the potential $V(\phi)$ between a unit charge and the neutralising background charge density $-\rho_{\rm b} = - N/2 \pi R$ \alert{as specified by (\ref{1.3})}, is independent of $\phi$. In the logarithmic case ($s=0$ in (\ref{1.11})) one calculates $V(\phi) = V_0 = N \log R$ by using the classical Jensen's formula. For the remaining \alert{permitted} values of the exponent $s$ in one-dimension
(the discussion below (\ref{1.11}) gives $-2<s<1$) use of a trigonometric form of the beta integral gives \alert{that the integral  (\ref{1.3}) has the evaluation}
\begin{equation}\label{2.1}
V(\phi) = V_0 = {\rm sgn} \, (s) N R^{-s} {\Gamma(1-s) \over (\Gamma(1-s/2))^2}.
\end{equation}
In terms of this $\phi$ independent potential, one calculates from (\ref{1.6}) and (\ref{1.6a}) that
\begin{equation}\label{2.2}
U_{\text{p-b}}+
U_{\text{b-b}} = N V_0/2.
\end{equation}
\alert{The essential point to make at this stage is in the thermodynamic limit, $N,R \to \infty$ with $\rho_{\rm b}$ fixed this diverges faster than $N$ for all permissible $s$.}

\alert{We would like to show that there is an equal and opposite leading divergence associated with the particle-particle potential energy in a typical configuration.  As an analytically tractable setting we specialise to the particular}
lattice configuration \alert{where} the mobile charges are at their minimum energy positions. \alert{O}n a circle this corresponds to equal spacing \cite{Ve78,BHS09}.\footnote{As a side note, on the two-dimensional sphere the problem of finding an energy-minimising configuration is known as Smale's 7th problem \cite{Sm00}; see also \cite{BHS19}, \cite{BT19} for a recent account.} For the logarithmic potential, the simple identity $\prod_{l=1}^N(z - e^{2 \pi i l/N}) = z^N - 1$, together with  (\ref{2.2}) give
$U_{\text{p-b}}+
U_{\text{b-b}}+U_{\text{p-p}} = - {1 \over 2} N \log 2 \pi \rho_{\rm b}$, exhibiting that it is linear $N$.
For $s \ne 0$ the sum giving the potential energy of the equally spaced configuration on the circle has been analysed in 
\cite[Th.~1.1 with an extra factor of $R^{-s}$ on the RHS to account for the circle having radius $R$]{BHS09}, with the result
\begin{equation}\label{2.2X}
U_{\text{p-p}} \sim -N V_0/2 + N {\rm sgn}\,(s) \rho_{\rm b}^s \zeta(s) (1 +
{\rm O}(N^{-2})),
\end{equation}
where $\zeta(s)$ denotes the Riemann zeta function.
Hence indeed the total potential energy $U_{\text{p-b}}
+U_{\text{b-b}}+U_{\text{p-p}}$ is extensive.
In relation to the logarithmic potential case, one notices that
${d \over ds} \rho_{\rm b}^s \zeta(s) |_{s=0} = - {1 \over 2} \log 2 \pi \rho_{\rm b} $, \alert{which is the correct free energy per particle. This prescription also allows the logarithmic potential to be deduced from the Riesz potential with $s > 0$ in (\ref{1.11}).}

Following the theme put forward in \cite[\S IV.A]{Le22}, the result so obtained for the leading order total energy of the Riesz gas on a circle of radius $R$, density $\rho_{\rm b} = N/2\pi R$
and exponent $0<s<1$
should be compared against that for the Riesz gas with the same density but exponent $s>1$. In the latter setting the Riesz potential is integrable at infinity, so one does without a neutralising background. For the particle-particle potential energy in the case of equal spacing we have
\begin{equation}\label{2.3}
U_{\text{p-p}} =
{N \over 2 R^s} \sum_{j=1}^{N-1}
{1 \over |1 - e^{2 \pi i j/N}|^s} \sim N \rho_{\rm b}^s \zeta(s), \quad s>1.
\end{equation}
\alert{Significantly this} is same result found for the leading total potential energy for $0<s<1$. Thus one can interpret the background as providing an analytic continuation of the leading total energy from the regime where no background is required.

We consider next the \alert{total} potential energy $E_N(s,\phi)$ say at an angle $\phi$ due to the particles fixed to be equally spaced (at $\theta = 2 \pi j / N$
with $j=1,\dots,N$ for definiteness), and the uniform background. 
\alert{Our aim here is to demonstrate that for $-2 < s < 1$
this is of order unity in the thermodynamic limit, and that it relates to the case $s>1$ without the background via an analytic continuation. }
\alert{This total potential energy is specified by} 
\begin{equation}\label{2.3a}
E_N(s,\phi) = \Big (\sum_{j=1}^N
\Psi_s(R e^{2 \pi i j/ N}, 
R e^{i \phi}) \Big ) - V(\phi).
\end{equation}
\alert{Excluding the case $s=0$ and setting $\rho_{\rm b}$, both for brevity}, 
by repeating the calculation of \cite[proof of Th.~1 to the first two orders]{BHS09}
for all $s$ in the range
 $-2<s<1$, $(s \ne 0$) gives
 \begin{equation}\label{2.3c}
 \lim_{N \to \infty} E_N(s,2 \pi x/N ) = \zeta(s;x) + \zeta(s;1-x), \quad 0<x<1,
 \end{equation}
 where $\zeta(s;a)$ denotes the Hurwitz zeta function.
 For $x \notin \mathbb Z$, the evaluation is deduced from this for general $x \in \mathbb R$ by the requirement of periodicity, period 1,  in $x$. \alert{In addition to this indeed being an order unity quantity, the further}
 point to be made here is that this is the analytic continuation in $s$ of the limiting large $N$ form of 
$E_N(s,2 \pi x/N )$ for $s>1$ with the term $V(\phi)$ deleted.
\alert{Thus from (\ref{2.3a})  and (\ref{1.11}), the latter is equal to the convergent sum $\sum_{n=-\infty}^\infty |n+x|^{-s}$,
which is a rewrite of the RHS of (\ref{2.3c}) for $s>1$.}

\subsection{Higher Dimensions}
For $d \ge 2$ we restrict our attention to
a discussion of the analytic properties of the \alert{total} potential energy at a point for lattice configurations
when a background is required, with our presentation consisting mostly of quoting results from \cite{Le22}.

Let $\{ \vec{r}_j \} \subset \mathbb R^d$ be such that $\inf_{j \ne k} |\vec{r}_j - \vec{r}_k| > 0$, and let $B_R(\vec{r})$ denote the ball in $\mathbb R^d$ of radius $R$ centred at $\vec{r}$.
In the $d$-dimensional case, it has been proved in \cite[Lemma 29 and associated text]{Le22} that
\begin{equation}\label{2.4}
\Phi(s,\vec{r}) :=
\lim_{R \to \infty} \Big (
\sum_{\{\vec{r}_j\} \in B_R(\vec{r})}
{1 \over | \vec{r} - \vec{r}_j|^s}
- \rho_{\rm b} \int_{B_R(\vec{r})}
{d \vec{r} \, '\over | \vec{r} - \vec{r} \, '|^s} \Big ) 
\end{equation}
in the range of exponents $d-2 < s < d$ ($d \ge 4$),
$d-2 < s < d-2+2/(d+1)$ ($d=2,3$)
is \alert{well defined and} an analytic continuation of the same quantity defined without the integral (background) term.

Specialising to a lattice $\mathcal L$
(integer span of a basis in $\mathbb R^d$), more can be said \cite[\S 2]{Le22}. 
Let us use the notation
$\Phi(s,\vec{r})$ for the corresponding lattice sum, which is well defined for $s> d$ and $\vec{r} \notin \mathcal L$. 
This a periodic function of $\vec{r}$, being invariant with respect to translations by the lattice vectors.
In the range $0<s<d$ the singularity at the lattice points --- and in particular at the origin --- is integrable, allowing for the construction of a (formal) Fourier series
\begin{equation}\label{2.5}
\sum_{\ell^* \in \mathcal L^*}
\alpha_{\vec{\ell^*}}
e^{2 \pi i \vec{r} \cdot \vec{\ell}}, \quad
\alpha_{\vec{\ell^*}} =
\int_{\mathbb R^d} | \vec{r}\,'|^{-s} e^{-2 \pi i \vec{\ell^*} \cdot  \vec{r}\,'} \, d \vec{r}\,',
\end{equation}
where $\mathcal L^*$ denotes the dual to the lattice $\mathcal L$. Why this is only a formal series is that the Fourier coefficient for $\vec{\ell^*} = \mathbf 0$ is divergent as the integral does not converge at infinity. Deleting this term, it can be proved \cite[Lemma 31]{Le22} that
(\ref{2.5}) provides the analytic continuation of 
$\Phi(s,\vec{r})$ in $s$ to the region $0<s<d$. It therefore coincides with the RHS of (\ref{2.4}) for the range of $s$ values noted below that equation, with
$\{\vec{r}_j\}$ therein being the lattice points of $\mathcal L$.
Moreover, if the domains of summation/ integration are appropriately modified to be consistent with the lattice structure, we have \cite[Lemma 31]{Le22} that the limit (\ref{2.4}) coincides with the analytic continuation, except in the Coulomb case $s=d-2$. In that circumstance, under the assumption that the Wigner-Seitz cell of $\mathcal L$ as zero quadrapole moment, the limit gives rise to an extra term
proportional to $|\vec{r}|^2$ integrated over the cell \cite[Eq.~(98)]{Le22}.

\section{Explicit Functional Forms of Electrostatic Potentials Due to a Uniform Background}\label{S3}

\alert{As revised, the one-component plasma model consists of a smeared out, uniform neutralising background inside a domain $\Omega$, in addition to the mobile point charges. A basic question then relates to the computation of the potential in this setting. Consideration of this problem for balls and hyperellipses leads to a quadratic potential within $\Omega$, with coefficients which are unchanged by dilation $\Omega \mapsto c \Omega$. This latter property ties in with Newton's shell theorem from the theory of the gravitational potential, implying that for these geometries the corresponding uniformly charged annular regions $\Omega \backslash c \Omega$ ($0<c<1$) are such that the electric field vanishes inside of the hollow $c \Omega $. In the case $d=2$ we review aspects of what may be termed the inverse problem in this setting, whereby the task is to compute $\Omega$ from knowledge of the potential energy created by the background.
For the uniformly charged disc, ellipse, and the annulus with $d=2$, we use knowledge of the explicit form of the potential to compute the energy $U_{\rm b-b} + U_{\rm p-b}$, which we show relates to the asymptotic expansion of the normalisation for certain random matrix ensembles.
}

\subsection{The Coulomb Potential with Ball Geometry}\label{S3.1}

We first consider the case that $\Omega$ is the (open) ball of radius $R$ in $\mathbb R^d$, centred at the origin. Denote this
$B_R$. We want to evaluate the \alert{corresponding} potential $V(\vec{r})$ as specified by (\ref{1.3}) with the background charge density the constant $C(\vec{r}) = - \rho_{\rm b}$. In this setting we know that up to additive harmonic functions, $V(\vec{r})$ has the functional form (\ref{1.4}), which is the solution of the Poisson equation $\nabla_{\vec{r}}^2 V(\vec{r}) =
 c_d \chi_d \rho_{\rm b} \mathbbm 1_{\vec{r} \in \Omega}$. The simplifying feature of $\Omega$ being
 a sphere is that $V(\vec{r})$ must be a function of only $r=|\vec{r}|$. We can then replace the operator $\nabla_{\vec{r}}^2$ by its radial component
 (we use the symbol $\doteq$ to indicate this) in spherical coordinates,
 \begin{equation}\label{3.1}
 \nabla_{\vec{r}}^2
 \doteq {1 \over r^{d-1}}
 {d \over d r} r^{d-1} {d \over dr}.
 \end{equation}
The Poisson equation can now be solved to find missing harmonic functions in (\ref{1.4}).

\begin{proposition}\label{P3.1}
For the ball geometry as specified above, with the ball filled by a uniform background charge density $-\rho_b$, total charge $-N$, the electrostatic potential $V(\vec{r}) = V(r)$ is given by
\begin{equation}\label{3.2}
V(\vec{r}) = 
\begin{cases}\displaystyle
\rho_{\rm b} \Big ({c_d \chi_d \over 2 d} \Big ) (r^2 - R^2) - N  \Phi_{\rm d} (R), &
 \vec{r} \in B_R, \\[.2cm]
- N  \Phi_{\rm d} (r), & {\rm otherwise},
\end{cases}
\end{equation}
where $\Phi_{\rm d} (r) = \Phi_{\rm d} (\vec{r},\vec{0})$ for any vector $\vec{r}$ such that
$\vec{r} = r$.
\end{proposition} 

\begin{proof}\footnote{\alert{The case $d=3$ is often discussed on the internet as a problem in elementary electrostatics, with its solution presented conditional on knowledge of Gauss' law, rather than based on (\ref{3.1}).}}
Consider first the potential outside of the ball. The RHS of the Poisson equation is then zero, and so restricting attention to radially symmetric solutions 
$V(\vec{r}) = V(r)$
as compatible with the symmetry of the ball, according to (\ref{3.1}) our task is to solve
\begin{equation}\label{3.3}
{d \over d r} r^{d-1} {d \over dr}
V(r) = 0.
\end{equation}
This elementary second order linear differential equation has general solution
\begin{equation}\label{3.4}
V(r) = 
\begin{cases}\displaystyle
- {A \over d - 2} r^{-d+2} + B, & d \ne 2,
\\
 A \log r + B, & d =2,
\end{cases}
\end{equation}
where the constants of integration $A,B$ may depend on $d$. To determine the latter, we take $|\vec{r}|$ large in (\ref{1.3}). This gives the asymptotic form  $-  \rho_{\rm b} \Phi_{\rm d}(\vec{r},\vec{0}) (1 +
{\rm O}(1/r))
\int_\Omega d \vec{r} = - N \Phi_{\rm d}(\vec{r},\vec{0}) (1 +
{\rm O}(1/r))
$, where to obtain the equality use has been made of (\ref{1.5}).
Recalling (\ref{1.2}) and the definition of $\chi_d$ noted above the latter, it follows $A = N \chi_d $, $B=0$. The second case of (\ref{3.2}) now follows.

For $\vec{r} \in B_R$ the RHS of (\ref{3.3}) is to be set equal to
$c_d \chi_d \rho_{\rm b}$. Making use of the first case of (\ref{1.4}) it follows that the missing additive harmonic function therein is a constant. Fixing its value by requiring continuity at the boundary of the ball with the outside solution completes (\ref{3.2}).
\end{proof}

By   (\ref{1.5}) the total charge in the ball due to the background is $-N$. One thus sees that the potential outside of the ball is the same as that for a point particle of charge $-N$ at the origin. This is (one of the) statements proved by Newton for $d=3$ in the context of the gravitational force, often referred to as the shell-point equivalent theorem (or simply Newton's shell theorem); for recent introductory articles relating to this, see
\cite{CR22}, \cite{Ca23}. The other statement of Newton's shell theorem relates to the potential due to the generalised annulus region $B_R \backslash c B_R$, $0<c<1$, in the hollow region $cB_R$.
Thus the fact that the coefficient of $r^2$ in (\ref{3.2}) is independent of $R$ tells us that the potential therein (obtained by subtracting from $V(r)$ the same functional form but with $R$ replaced by $cR$) will be equal to a constant in $cB_R$, and thus the electric field vanishes.

This example also draws attention to another effect in potential theory, namely the possibility of a mother body, whereby a measure supported inside the bounding domain  of the charge distribution produces the same potential as the domain itself (at least outside the bounding domain).  Historical aspects of this topic can be found in 
e.g.~\cite[Introduction section]{BS16}.

There is an application of Proposition \ref{P3.1} in the case
$\vec{r} \in B_R$, $d=2$ to random matrix theory (see \S \ref{S3.4}).
In preparation, we so specialise the $d$-dependent quantities therein to read off that then
\begin{equation}\label{3.4a}
V(\vec{r}) = {\pi \rho_{\rm b} \over 2} r^2 - {N \over 2} + N
\log R.
\end{equation}
With this established, we make use of (\ref{1.6}) and (\ref{1.6a}) to calculate that in this setting
\begin{equation}\label{3.4b}
U_{\text{b-b}} + U_{\text{p-b}} = {\pi \rho_{\rm b} \over 2}
\sum_{j=1}^N |\vec{r}_j |^2 - {3N^2 \over 8} + N^2  \log R, \quad r\le R.
\end{equation}
As a variation also relevant to random matrix theory, we make note of the potential in the \alert{annulus} region $cR\le r \le R$ \alert{with} uniform charge density 
$-\rho_{\rm b} = - N/(\pi(1-c^2)R^2)$.
Following the subtraction procedure in the case $d=2$ of (\ref{3.2}) (with $N$ replaced by $\rho_b |B_R|$) as noted in the paragraph two above, we calculate
\begin{equation}\label{3.4ai}
V(\vec{r}) = {\pi \rho_{\rm b} \over 2} (r^2 - R^2)  + \pi R^2 \rho_{\rm b} (\log R - c^2 \log r),
\end{equation}
and 
\begin{align}
\begin{split}
\label{3.4bi}
 U_{\text{b-b}} + U_{\text{p-b}} 
&= {\pi \rho_{\rm b} \over 2}
\sum_{j=1}^N |\vec{r}_j |^2 - {\pi \rho_{\rm b} N R^2 (3 + c^2) \over 8} 
\\
&\quad + {\pi \rho_{\rm b} N R^2 \over 2} (\log R - c^2 \log r), \quad c R \le r\le R.
\end{split}
\end{align} 

There is also an application of Proposition \ref{P3.1} 
in the case
$\vec{r} \in B_R$, $d=1$ to a vicious walker problem from statistical mechanics \cite{Fi84},
which is to be discussed in 
\S \ref{S3.5.1} below.
In this case
\begin{equation}\label{3.4d}
V(x) =  \rho_{\rm b}  (x^2 - R^2)
+   N R,
\end{equation}
from which (\ref{1.6}) and (\ref{1.6a}) give
\begin{equation}\label{3.4e}
U_{\text{b-b}} + U_{\text{p-b}} =
\rho_{\rm b} \sum_{j=1}^N x_j^2 + {N^2 R \over 6}.
\end{equation}

\subsection{Coulomb Potential with Hyperellipsoid Geometry}

We suppose now that $\Omega$ is the interior of the hyperellipsoid surface
\begin{equation}\label{3.5}
{x_1^2 \over a_1^2} + \cdots +
{x_d^2 \over a_d^2} = 1,
\end{equation}
with $a_1,\dots,a_d > 0$; the case that these latter parameters are all equal to $R$ then corresponds to the ball $B_R$ in dimension $d$. However unlike the situation with $B_R$ in the Coulomb case, there is no symmetry providing an effective dimensional reduction as follows from (\ref{3.1}). Instead, following a \alert{refinement from \cite{Di16}}
to a classical work of Dirichlet \cite{Di39}, \alert{the approach to be taken} is to work directly with the multi-dimensional integral (\ref{1.3}).

\begin{proposition}\label{P3.2}
For $d > 2$, consider (\ref{1.3}) with the Coulomb potential (\ref{1.2}), and \alert{constant} background charge density
$C(\vec{r}\,') = - \rho_{\rm b}$ supported inside of the domain $\Omega$, with total charge equallying $-N$.
Set 
 \begin{equation}\label{3.12}  
 \quad S_0(\lambda) = (a_1^2 + \lambda) \cdots (a_d^2 + \lambda), \quad
 S_1(\lambda) = {x_1^2 \over a_1^2 + \lambda} + \cdots +
 {x_d^2 \over a_d^2 + \lambda}.
 \end{equation}
 We have
 \begin{equation}\label{3.13} 
 V(\vec{r}) =
  - N d (d/2 - 1)
 \int_0^\infty   {1
 \over \sqrt{S_0(\lambda})}
 (1 - S_1(\lambda)) \mathbbm 1_{S_1(\lambda) < 1} \,  d \lambda.
  \end{equation}
\end{proposition}

\begin{proof}
For $k \in \mathbb R_{\ge 0}$, use will be made of the Fourier transform
\begin{equation}\label{3.6}
{1 \over \pi}
\int_{-\infty}^\infty {\sin y \over y} e^{i k y} \, dy =
{1 \over 2 \pi i} 
\int_{C_\epsilon}
\Big ( {e^{i y (k+1)} \over y} -
 {e^{i y (k-1)} \over y} \Big ) \, dy =
 \begin{cases}1, & 0 \le k < 1, 
 \\
 0, & k>1.
 \end{cases}
 \end{equation}
 Here $C_\epsilon$ is the real line with the segment $[-\epsilon, \epsilon]$ replaced by the \alert{complex} half circle indentation $\epsilon e^{i \theta}$, with $\theta$ starting at $\pi$ and finishing at $0$. We use (\ref{3.6}) with
 $k$ \alert{equal to the LHS of (\ref{3.5})}
 as an indicator function for the interior of the hyperellipsoid surface. Noting that
 \begin{equation}\label{3.8}
 {1 \over | \vec{r} - \vec{r} \, ' |^s} =
 {\xi_s  \over \Gamma(s/2)} 
 \int_0^\infty v^{s/2 - 1} e^{i v | \vec{r} - \vec{r} \, '|^2} \, dv, \quad 2 > {\rm Re}(s)> 0,
  \end{equation}
 \alert{for a particular $\xi_s$ of unit modulus},  it follows upon changing the order of the $v$ and $y$ integrations that
 \begin{align}
\begin{split}
    \label{3.9} 
 V(\vec{r}) & =
 - {\xi_s \rho_{\rm b} \over 2 \pi i \Gamma(s/2)} 
 \int_0^\infty d v \,  v^{s/2 - 1} \int_{C_\epsilon} {dy \over y} \\
 &\quad  \times \int_{\mathbb R^d} d \vec{r} \, '\,
 \Big ( e^{i y} e^{i y k + i v |\vec{r} - \vec{r} \, '|^2} -
e^{-i y} e^{i y k + i v |\vec{r} - \vec{r}\, '|^2} \Big ).
 \end{split}
 \end{align}

  Recalling the definition  of $k$ as the LHS of (\ref{3.5}), we see that the multidimensional integral over $\vec{r} \in \mathbb R^d$ separates into a product of $d$ univariate integrals of the form
\begin{equation}\label{3.10} 
  \int_{-\infty}^\infty
  e^{i x^2 y/a_j^2 + i v(x - x_j)^2} \, dx = a_j \sqrt{\pi} 
  {e^{i \pi/4} \over \sqrt{y + a_j^2 v} }
  e^{i v x_j^2 y / (y + a_j^2 v)}.
\end{equation}
Substituting in (\ref{3.9}) and changing variables
$v = y/\lambda$ for fixed $y$ then shows
\begin{align}
\begin{split}
\label{3.11} 
 V(\vec{r}) & =
 - {\xi_s \rho_{\rm b} A_d\over 2 \pi i \Gamma(s/2)} 
 \int_0^\infty d \lambda \,  {\lambda^{(d-s)/2 - 1} 
 \over \sqrt{S_0(\lambda})}
 \\
 &\quad \times \int_{C_\epsilon} dy \, y^{(s-d)/2-1} (e^{iy(1 + S_1(\lambda))}
 - e^{-iy(1-  S_1(\lambda))} ),
\end{split}
\end{align} 
  where $\xi_s$ again has the property that $|\xi_s| = 1$, but
  may be different to that in (\ref{3.8}), \alert{while}
  $ A_d = \pi^{d/2}  \prod_{j=1}^d a_j$ and the functions $S_0(\lambda)$,
  $S_1(\lambda)$ are as in (\ref{3.12}). In arriving at (\ref{3.11}),
  deformation of the integration domain for $\lambda$
  has been carried out in the cases that $y \notin \mathbb R^+$.

A feature of this expression is that it is  well defined
for $s$ in the range $0 < {\rm Re}(s) < d$, thus extending the
earlier restriction from (\ref{3.8}).
Specialising now to $s=d-2$ $(d > 2)$, the integral over $y$ can
can be evaluated by closing the contour $C_\epsilon$ in the lower half plane, showing that after dividing by $2\pi i$ it evaluates to
$-i (1 - S_1(\lambda)) \mathbbm 1_{S_1(\lambda) < 1}$. The result
(\ref{3.13}) now follows  upon noting that $|\Omega| = A_d/\Gamma(1+d/2)$, use of (\ref{1.5}) \alert{and determining
the unimodular factor $\xi_s$  by consideration of the ball case $a_1=
\cdots = a_d$ and comparison with (\ref{3.2})}.
\end{proof}

A feature of (\ref{3.13}) is that for $\vec{r} \in \Omega$, the factor
$ \mathbbm 1_{S_1(\lambda) < 1}$ inside the integrand is unity for
all $\lambda \in [0,\infty)$. Consequently in this circumstance the potential is a quadratic polynomial
 \begin{equation}\label{3.13a} 
 V(\vec{r}) = \sum_{j=1}^d \alpha_j x_j^2  + \alpha_0,
 \end{equation}
where the constants, which depend on $\{a_j\}$ \alert{in (\ref{3.5}),} are specific definite integrals which can be read off from (\ref{3.13}). Moreover, one can check from
(\ref{3.13}) that scaling the \alert{hyper}ellipsoid $a_j \mapsto c a_j$, $c>0$,
leaves $\{\alpha_j\}$ in (\ref{3.13a}) unchanged. Consequently, if one was to consider the potential due to $\Omega \backslash c \Omega$ with $0<c<1$, as for the case that $\Omega = B_R$ and
Newton's shell theorem (recall the paragraph below the proof of Proposition \ref{P3.1}), this will equal a constant in the hollow
region $c \Omega$ (see also \cite{KL14}).

One notes that (\ref{3.13a}) is consistent with
(\ref{1.4}), which moreover implies the sum rule
 \begin{equation}\label{3.13b}
\sum_{j=1}^d \alpha_j = \rho_{\rm b} {c_d \chi_d \over 2}.
  \end{equation}
 Another point of interest is 
  that it is known (see e.g.~\cite[Remark 1.1 and references therein]{DF16}) that the hyperellipsoid provides the most general compact domain such the potential is a quadractic function of the components of the position vectors in the interior.
   Also noted in \cite[\S 3]{DF16} is that for $d=3$ the $\alpha_j$ (referred to as demagnetising factors) can be expressed in terms of certain incomplete elliptic integrals of the second kind.
For $|\vec{r}| \to \infty$, the condition $S_1(\lambda) < 1$ in (\ref{3.13a}) simplifies to $\lambda > |\vec{r}|^2$, since then $S_1(\lambda)$ can be replaced by $|\vec{r}|^2/\lambda$. Now that we know $\lambda$ must be large, we can replace $S_0(\lambda)$ by $\lambda^d$. This shows $V(\vec{r}) \sim - N \Phi_d(\vec{r}, \vec{0})$ as must be from the definition (\ref{1.3}).

For $d=2$ the integral in (\ref{3.13}) diverges due to the integrand not being integrable at infinity, while the prefactor $d/2-1$ vanishes. 
In fact the $d=2$ case can be treated separately. 

\begin{proposition}
    For points inside a uniformly charged ellipse
   \begin{equation}\label{3.13c} 
 V(\vec{r}) = {\pi \rho_{\rm b} \over 2} \Big (
 |\vec{r}|^2 -  {a_1 - a_2 \over a_1 + a_2} (x^2 - y^2)  + 2 a_1 a_2 
 \log{a_1+a_2 \over 2} - {a_1 a_2} \Big ).
 \end{equation}
 Consequently
 \begin{equation}\label{3.4f}
U_{\rm{b}\text{-} \rm{b}} + U_{\rm{p}\text{-}\rm{b}} = {\pi \rho_{\rm b} \over 2}
\sum_{j=1}^N \Big ( |\vec{r}_j |^2 
-  {a_1 - a_2 \over a_1 + a_2} (x_j^2 - y_j^2) \Big )
- {3N^2 \over 8} + N^2 \log {a_1 + a_2 \over 2};
\end{equation}
cf.~(\ref{3.4b}).
\end{proposition}

\begin{proof}  \alert{
   Up to the constant term, (\ref{3.13c}) is a classical result due to Lamb \cite[\S 159]{La32}, derived independently in
  \cite{CPR87}. The computation of the constant terms has been
  provided in \cite{VY10}. 
  We provide an independent derivation of (\ref{3.13c}) in
  Proposition \ref{P3.3a} below.
  With (\ref{3.13c}) established, 
  (\ref{3.4f}) follows from the definitions and straightforward integration.}
\end{proof}

As with (\ref{3.4b}), we highlight this two-dimensional case for its application to random matrix theory (see \S \ref{S3.4}).
  Furthermore, as noted in the case of the hyperellipsoids $d > 2$, one observes that \alert{(\ref{3.13c})} is consistent with (\ref{1.4}), and moreover that the quadratic terms are invariant under scaling of the ellipse $\Omega \mapsto c \Omega$.

\subsection{Coulomb Potential in the Case of a Rectangular Cuboid and Rectangle}

For the cuboid $\Omega = [a_1,b_1] \times [a_2,b_2] \times [a_3,b_3]$, one learns from \cite[\S 1(c)]{Bo22} that the corresponding electrostatic potential (with $\rho_{\rm b} = 1$)
\begin{equation}\label{3.15}
V(y_1,y_2,y_3) = \int_\Omega {1 \over \sqrt{(y_1-x_1)^2 + (y_2-x_2)^2+(y_3-x_3)^2}} \, dx_1 dx_2 dx_3,
\end{equation}
was first considered in a 1930 monograph by MacMillan \cite[pp.~72-80]{Ma30}. In this work, a closed form evaluation involving only the elementary functions artanh and arctan was given, albeit involving 48 terms and a page and a half to present. (For the case of a cube, see also the expressions obtained in
\cite{Hu96} and \cite{Ci15}.) One learns too from
\cite[\S 1(c)]{Bo22} that with $\rho_\delta = \sqrt{\delta_1^2+\delta_2^2+\delta_3^2}$ and $\sum_{{\rm cyc.}\, \delta}$
denoting the cyclic sum in $(\delta_1,\delta_2,\delta_3)$, 
the long expression from \cite{Ma30} can be cast in the very efficient and symmetric form \cite{Wa76}
\begin{equation}\label{3.16}
V(y_1,y_2,y_3) = 
\sum_{\substack{\delta_j \in \{y_j-a_j, b_j - y_j\} \\ (j=1,2,3)}}
\sum_{{\rm cyc.}\, \delta}\Big (
\delta_1 \delta_2 \, {\rm arctanh}\, {\delta_3 \over \rho_\delta} -
{\delta_1^2 \over 2} \arctan {\delta_2 \delta_3 \over \delta_1 \rho_\delta} \Big ).
\end{equation}
Also in \cite[\S 1(c)]{Bo22}, attention is drawn to the closed form expression for the background electrostatic energy in the case of the unit cube ($\rho_{\rm b} =1$ for convenience) from 
\cite{Wa76}
\begin{equation}\label{3.17}
U_{\text{b-b}} =
{1 \over 5}\Big ( 1 + \sqrt{2} - 2 \sqrt{3} \Big )
+ \log \Big ( (1 + \sqrt{2})(2 + \sqrt{3}) \Big ) - {\pi \over 3};
\end{equation}
see also \cite{Ha02}, \cite{Ci11}.

In the case of a rectangle $[a_1,b_1] \times [a_2,b_2]$, generalising the square case from \cite{Hu96}
with the logarithmic potential,
one has for the potential $V(x,y)$ due to a uniform background (set equal to unity)
\begin{equation}\label{3.18}
V(x,y) = - {1 \over 2}
\sum_{i,j=0}^1 \bigg (
\alpha_i \beta_j \log (\alpha_i^2 + \beta_j^2) -
3 \alpha_i \beta_j + \beta_j^2
\arctan {\alpha_i \over y_j} +\alpha_i^2
\arctan {\beta_j \over \alpha_i} \bigg ),
\end{equation}
with $\alpha_0 = a_1 - x$, $\alpha_1 = b_1 -x$,
$\beta_0 = a_2 - y$, $\beta_1 = b_2 -y$.

\subsection{The Domain $\Omega$ as Implied by the Potential}\label{S3.4A}

\alert{In this subsection we restrict to $d=2$}, and begin by
considering the potential (\ref{3.4ai}). Since $\log r$ is harmonic in its domain of validity
(the annulus $cR \le r \le R$), we know from Poisson's equation as noted above (\ref{3.1}) that the corresponding background charge density must be equal to $-\rho_{\rm b}$. We pose the question as to what features of the functional form (\ref{3.4ai}) distinguishes the boundary of the annulus? Consider first the inner radius. We see that its value $r=cR$ can be distinguished as the unique solution of the equation
$r {d \over dr} V(r) = 0$ (here we have included as extra factor of $r$ to allow for the case $c=0$).
With this noted, we can make use of the normalisation condition implied by the Poisson equation
$\int_\Omega \nabla_{\vec{r}}^2 V(r) \, d \vec{r} =
2 \pi \rho_{\rm b} | \Omega|$ to see that the outer boundary $r=R$ can be distinguished by the requirement that $r {d \over dr} V(r) = \rho_{\rm b} | \Omega|$.
These findings are consistent with the following more general result (identify $q(r) = 2 V(r)$) \cite[\S IV.6]{ST97}.

\begin{proposition}
Let $q(r)$ be strictly subharmonic, meaning that its Laplacian is strictly positive.
Consider the  two-dimensional radially symmetric log-potential Coulomb system with Boltzmann factor proportional to
\begin{equation}\label{3.19}
e^{- {\beta N \over 2} \sum_{j=1}^N q(|\vec{r}_l|)}
\prod_{1 \le j < k \le N} |
\vec{r}_k - \vec{r}_j|^{\beta}.
\end{equation}
For $N \to \infty$ the support of the limiting density (in the present context often referred to as the droplet) is the annulus $r_0 \le r \le r_1$ with $r_0, r_1$ the solutions of
\begin{equation}\label{3.20}
r {d \over dr} q(r)=0, \qquad r {d \over dr}q(r) = 2
\end{equation}
respectively.
    \end{proposition}

    Attracting a lot of attention for its relevance to certain fluid evolutions in the plane
    \cite{Ri72}, \cite{Za06}, \cite{GTV14}, \cite{LM16}, \cite{RM25} is the question of the support of the equilibrium for (\ref{3.19}) with $q(|z|)$ replaced by $|z|^2 + 2 {\rm Re} \, P(z)$ where ${\rm Re} \, P(z)$ is harmonic (here $\mathbb R^2$ is identified by $\mathbb C$.) Note that the potential (\ref{3.13c}) for  the one-component plasma confined to an ellipse can be written in this form (up to a constant factor), with $P(z)$ proportional to $z^2$. For general $P(z)$, from Poisson's equation the implied background density is the constant
    $\rho_{\rm b} = N/\pi$ (for convenience we will set $N=1$), with the task then being to compute the boundary of the support. This modified potential must satisfy
    \begin{align}
    \begin{split}
    \label{3.21}  
 &\quad {1 \over 2} \Big ( |z|^2 + 2 {\rm Re} \, P(z) \Big ) + C = 
 {1 \over \pi} \int_{\Omega} \log |z - w| \, d^2 w 
 \\
 &\implies \bar{z} + P'({z}) = {1 \over  \pi} \int_{\Omega} {1 \over  z - w} \, d^2 w, \: \:  z \in \Omega,
    \end{split}
    \end{align} 
 where $C$ is some constant. Here the second equation follows from the first by applying the derivative operator $\partial_{z}$ to both sides. 
 
 Consider now the analytic function defined for $z \in \mathbb C \backslash \bar{\Omega}$
 by the RHS of the second equation of (\ref{3.21}). It has the large $|z|$ leading form $|\Omega|/(\pi z)$.
Use of the Cauchy-Pompeiu formula allows the rewrite of the second equation in (\ref{3.21}),
 $
 P'(z) = {1 \over 2 \pi i}
 \int_{\partial \Omega}
 {\bar{w} \over z - w} \, d^2 w$.
 Key in relation to this is that it implies the boundary value of the Cauchy transform $H_\Omega(w) = 
 \bar{w} + P'(w)$ for $w \in \partial \Omega$ \cite{Ri72},
 \alert{where $H_\Omega(w)$ is defined as the RHS of the second equation in (\ref{3.21}) with $z$ outside of $\Omega$}. The idea from here
  is to introduce the conformal map $\xi(z)$, chosen to be proportional to $z$ for large $|z|$ with a positive proportionality, from the exterior of the unit disk to the exterior of $\Omega$. Assuming this to be real for $z$ real, we must then have \cite{EE97}, \cite{EM07}
\begin{equation}\label{3.22}  
H_\Omega(\xi(u)) =
\xi(1/u) + P'(\xi(u)), \quad |u| \ge 1.
 \end{equation}
 
 The fact that $H_\Omega(w)$ must be analytic for $|u| > 1$, with a specific decay at infinity, can now be used to specify $\xi(u)$, subject to an ansatz for the latter. As an example, suppose $P(z) = - \alpha z^2$ and make the ansatz $\xi(u) = c_1 u + c_0 + c_{-1}/u$. Consideration of the corresponding large $u$ behaviour of both sides of (\ref{3.22}) implies $c_1^2 = |\Omega|/(\pi(1 - (2 \alpha)^2)$, $c_0 = 0$, $c_{-1} =  2 \alpha c_1$. Also, making use of the complex form of Green's theorem, and then parametrising $\partial \Omega$ in terms of $\xi(u)$ allows for a residue evaluation of the volume,
 $|\Omega| = \pi(c_1^2 - c_{-1}^2)$.
 We can check that for
 $0 < 2 \alpha < 1$
 these findings characterise 
 $\Omega$ as an ellipse consistent with (\ref{3.13c}) (in the first sentence in Proposition \ref{P4.2} below, $\xi(u)$ identified as the Joukowski map).
 Moreover, these same methods can be used to calculate the constant $C$ in (\ref{3.21}).

 \begin{proposition}\label{P3.3a}
     \alert{With $P(z) = - \alpha z^2$, and $\alpha = 2 (a_1 - a_2)/(a_1+a_2)$, the first equation in
     (\ref{3.21}) is valid for $\Omega$ the ellipse with
     boundary $(x_1/a_1)^2+(x_2/a_2)^2=1$ and
     \begin{equation}\label{W19}
     C = 2 a_1 a_2 \log {a_1 + a_2 \over 2} - a_1 a_2.
     \end{equation}}
 \end{proposition} 
 \begin{proof}
 \alert{
Let $W(Z) = {1 \over \pi} \int_\Omega \log | Z - z| \, d^2z$.
From the meaning of $H_\Omega$ in (\ref{3.22}) as the Cauchy transform, after substituting $P(z) = - \alpha z^2$, multiplying both sides  by $\xi'(u)$, integrating from $1$ to $u_1$ with respect to $u$ and simplifying using the explicit form of $\xi(u)$
as deduced above we deduce
 \begin{equation}\label{W11}
W(\xi(u_1)) - W(\xi(1)) = {|\Omega| \over \pi}
\int_1^{u_1} {1 \over u} \Big ( 1 - {2 \alpha \over u^2}
\Big ) \, du \: \: \implies 
W(\xi(1)) = {|\Omega| \over \pi}  ( \log c_1 + \alpha).
 \end{equation}
Here the second equation follows upon evaluating the integral on the RHS and taking $u_1 \to \infty$ using  the fact that for $u_1$ large
$\xi(u_1) \sim c_1 u_1$ and so $W(\xi(u_1)) \sim  {|\Omega| \over \pi}   \log (c_1 u_1)$. On the other hand, from the first equation in (\ref{3.21}) with $z = \xi(1)$ we deduce
$W(\xi(1)) = {|\Omega| \over 2 \pi}(1 + 2 \alpha) + C$. Substituting this in the second equation of (\ref{W11}) we arrive at (\ref{W19}).
}
\end{proof}
 
 A \alert{further} example \alert{of this method} is given in the recent work
 \cite{BFL25}, where $P(z)$ is proportional to a particular logarithm --- see \cite{BBLM15} for the first work along these lines, as well as the subsequent contributions \cite{BM15}, \cite{ABK21}, \cite{CK22}, \cite{By24}, \cite{BY25}.

\subsection{Applications}\label{S3.4}

\subsubsection{Non-Intersecting Brownian Particles on a Line}\label{S3.5.1}
We begin by considering not non-intersecting Brownian particles, but rather a classical gas of particles $\{x_j\}$ on the real line with total energy
\begin{equation}\label{E.1}
U= {c \over 2} \sum_{j=1}^N x_j^2 - \sum_{1 \le j < k \le N} \log\Big | \sinh {\pi (x_k - x_j) \over L} \Big |.
\end{equation}
One sees that for large separations $|x_k - x_j| \to \infty$ the sum over pairs has the leading order form $-(\pi/L) \sum_{1 \le j < k \le N}
|x_k - x_j|,
$ and thus is proportional to the particle-particle potential energy for a one-component one-dimensional Coulomb system. Moreover, suppose that the one-dimensional Coulomb system has a uniform background in $[-R,R]$ with  $R = \pi N /c L$.
Then according to (\ref{3.4e}) the quadratic term in (\ref{E.1}) can be interpreted as resulting from the interaction of the particles with this uniform background. With the particle density of a one-component Coulomb system in equilibrium matching that of the background (if not, there would be an electric field and the system would be out of equilibrium \alert{as previously commented}), it follows from these Coulomb gas considerations that for large $N$ the particle density for the gas given by (\ref{E.1}) will be uniform in the interval 
$[-\pi N/c L, \pi N/c L]$, and zero outside of this interval. We know from (\ref{3.2}) that outside of the background, the Coulomb potential in the one-dimensional case is $N |x|$ and thus the one-body term in (\ref{E.1}) no longer matches. However the particles are outside of the region of the background  with low probability, so this has no effect on our conclusion.

Predictions can also be made for the large $N$ form of the configuration integral
\begin{equation}\label{E.2}
Q_{N,\beta} :=
\int_{-\infty}^\infty dx_1 \cdots 
\int_{-\infty}^\infty dx_N
\,
e^{-\beta (c/2) \sum_{j=1}^N x_j^2}
\prod_{1 \le j < k \le N}
 \Big | 2 \sinh {\pi (x_k - x_j) \over L} \Big |^\beta.
\end{equation}
From the viewpoint of a one-component Coulomb system, 
missing from the implied Boltzmann factor in (\ref{E.2}) are the terms independent of $\{x_j\}$ in
$e^{-\beta \pi (U_{\text{b-b}} + U_{\text{p-b}})/L}$.
\alert{Here} the factor of $\pi/L$ is due to this multiplying the effective one-dimensional Coulomb potential; recall the text below (\ref{E.1}).
\alert{Making use of} (\ref{1.8}), \alert{it follows that the missing factor} is equal to
$\exp(-\beta \pi N^2 R/6L) =
\exp(-\pi^2 \beta N^3/6cL^2)
$. The existence of the thermodynamic limit for a one-component Coulomb system as discussed below
(\ref{1.7}) then gives that for large $N$, with $c,L$ fixed, $- \pi^2 \beta N^3/6cL^2 + \log {1 \over N!} Q_{N,\beta}$ is of order $N$. We therefore conclude that the large $N$  expansion of $\log Q_{N,\beta}$ is $\pi^2 \beta N^3/6cL^2 + N \log N$, up to terms of order $N$. This can be checked in the case $\beta = 2$, when we have the exact evaluation \cite{Ti04}
\begin{equation}\label{E.3}
Q_{N,\beta} \Big |_{\beta = 2} =
 (\pi/c)^{N/2} N!
e^{g N (N^2 - 1)/6} \prod_{j=1}^{N-1} (1 - q^j)^{N-j}, 
\end{equation}
where $g = 2\pi^2/cL^2$ and
$q = e^{-g}$ --- the predicted large $N$ form of the logarithm is immediate. 

Following the presentation in \cite{Fo22}, we now link up the configuration integral (\ref{E.2}) with the non-intersecting Brownian particles problem. For diffusion of a single particle on the line, let $u_t(x^{(0)};x)$ denote the PDF for the event that starting at $x^{(0)}$ the particle arrives at $x$. For $N$ Brownian particles starting at $\mathbf x^{(0)} = (x_1^{(0)},\dots,x_N^{(0)})$ and finishing at $\mathbf x = (x_1,\dots,x_N)$ under the condition that they do not intersect, the corresponding PDF $G_t(\mathbf x^{(0)};\mathbf x)$ is given in terms of the single particle PDF by the determinant (Karlin-MacGregor) formula
$G_t(\mathbf x^{(0)};\mathbf x) = \det [
u_t(x_j^{(0)};x_k)
]_{j,k=1}^N $; see e.g. \cite[\S~4.3.1]{HKPV08}. Choosing the equal spacing initial condition $x_j^{(0)} = a(-(N+1)/2 + j)$ $(j=1,\dots,N)$ allows $G_t(\mathbf x^{(0)};\mathbf x)$ to be expressed in the form of the integrand of (\ref{E.2}) with ${\pi \over L} = {a \over 2 D t}$, $c = {1 \over D t}$, $\beta = 1$, and multiplied by
\begin{equation}\label{E.4}
(2 \pi D t)^{-N/2}
e^{a^2 \sum_{j=1}^N ((N-1)/2)^2 - (j-1)^2)/2 Dt}. 
\end{equation}
The probability that the particles do not intersect up to time $t$ is given by integrating over $\mathbf x$, and so is equal to $Q_{N,\beta}
|_{\beta = 1}$ with $c,L$ as identified, times (\ref{E.4}).
The predicted large $N$ form of $\log Q_{N,\beta}|_{\beta = 1}$  as noted above (\ref{E.3}) shows that the leading $N^3$ behaviour implied in the exponent of (\ref{E.4}) cancels out, giving that the implied order $N^2$ term from the exponent  is the leading form of this probability
\cite{Fo89}.

\subsubsection{Complex Ginibre Random Matrix Ensembles}\label{S3.4.2}
Let $X$ be an $N \times N$ standard complex Gaussian random matrix. A now classical result in random matrix theory due to Ginibre \cite{Gi65}
(see too the recent text
\cite[Prop.~2.1]{BF25}) is that the joint eigenvalue PDF has the explicit functional form
\begin{equation}\label{E.5}
{1 \over C_N}
\prod_{l=1}^N e^{-|z_l|^2}
\prod_{1 \le j < k \le N}
|z_k - z_j|^2, \qquad C_N = \pi^N \prod_{j=0}^{ N-1 } j!.
\end{equation}
Here the eigenvalues are complex valued. Each $z_l=x_l + i y_l$ can \alert{also} be \alert{regarded} as a point in the plane $\vec{r}_l = (x_l,y_l)$. Then $|z_k - z_j| = |\vec{r}_k - \vec{r}_j|$, $|z_l|^2 =
|\vec{r}_l|^2$, revealing that (\ref{E.5}) can be viewed as being proportional to the Boltzmann factor for a one-component Coulomb system with total energy (up to constant terms)
$
U = {1 \over 2} \sum_{j=1}^N |\vec{r}_j|^2 -
\sum_{1 \le j < k \le N} \log |\vec{r}_k - \vec{r}_j|
$
and with dimensionless inverse temperature $\beta = 2$; cf.~(\ref{1.9}). This analogy was already made by Ginibre \cite{Gi65}.

From the Coulomb gas viewpoint, the quadratic term in (\ref{E.5}) can be recognised from Proposition \ref{P3.1} as resulting from a uniform smeared out background charge density $\rho_{\rm b} = - {1 \over \pi}$ in  a disk of radius $R = \sqrt{N}$. The prediction is then that for large $N$ the eigenvalue density will equal ${1 \over \pi} \mathbbm 1_{|z| < \sqrt{N}}$, \alert{which is indeed in agreement with} explicit calculation \cite{Gi65}. This functional form is known in random matrix theory as the circular law, and in fact holds true in the general setting of random matrices with independent entries of zero mean and unit standard deviation, without further assumptions; see e.g.~\cite{BC12}. Also, the constant terms in (\ref{3.4}), together with knowledge of the free energy for the one-component plasma being extensive in the large $N$ limit \alert{gives the prediction} that $\log {1 \over N!} Z_N \sim \frac12 N^2 \log N - {3 \over 4} N^2 + {\rm O}(N)$. \alert{Here} $Z_N$ denotes the configuration integral over $z_l \in \mathbb C$ ($l=1,\dots,N$) of the functional form in (\ref{E.5}) without the normalisation constant. Knowledge of the normalisation constant allows this prediction to be verified.

Let $G$ be from a complex Ginibre ensemble (GinUE) and construct the matrix 
$
Y = {1 + \sqrt{c} \over 2} G + {1 - \sqrt{c} \over 2} G^\dagger$, $c = {1 - \tau \over 1 + \tau}$, $(0 \le \tau < 1)
$. The case $c=1$ is then the original GinUE, while $c=0$ gives the Hermitian random matrix ensemble (Gaussian unitary ensemble) noted above (\ref{1.8}). Introduce a scaled version of $Y$ by setting $\tilde{Y} = \sqrt{1 + \tau}
Y$.
Up to normalisation, the corresponding eigenvalue PDF  was analysed  in \cite{FKS97},
while the normalisation in the current conventions can be read off from the more recent work \cite[Eq.~(1.12)]{ABO25}.
It follows that
the only modification to the eigenvalue PDF relative the original GinUE occurs in the one-body factors (and the normalisation),
requiring that 
\begin{align}
\begin{split}
\label{p.s}
&e^{- \sum_{j=1}^N | z_j|^2} \mapsto \exp \Big ( - {1 \over 1 - \tau^2} \sum_{j=1}^N \Big ( |z_j|^2 - {\tau \over 2} (z_j^2 +
\bar{z}_j^2 ) \Big ) \Big ), 
\\
& C_N \mapsto
\pi^N (1 - \tau^2)^{N/2}  \prod_{j=0 }^{ N-1 } j!.
\end{split}
\end{align} 
As noted in \cite{DGIL94},
\cite{FJ96}, this can be viewed as a one-component Coulomb gas in the plane with logarithmic pair potential as for the complex Ginibre case. However now the one body potential    results from a uniformly
charged ellipse, charge density $-1/\pi (1 - \tau^2)$ supported inside of the ellipse with semi-axes $A = \sqrt{N}(1 + \tau)$,
$B = \sqrt{N}(1 - \tau)$;
this can be seen by comparing with
(\ref{3.4f}).
We remark that this is consistent with the elliptic law in random matrix theory
(see e.g.~\cite{NO14}) as generalising the circular law. It explains too why the corresponding random matrix ensemble is referred to as elliptic GinUE \cite[\S~2.3]{BF25}.

We turn our attention now to the asymptotic expansion of $\log {1 \over N!} Z_N$, where $Z_N$ the configuration 
integral previously considered in the complex Ginibre case, 
but with the first replacement in (\ref{p.s}). This is 
predicted from (\ref{3.4f}) to be
$\log {1 \over N!} Z_N \sim  N^2 \log {A+B \over 2} - {3 \over 4} N^2 + {\rm O}(N)$, which we see agrees with that implied by the modified normalisation in (\ref{p.s}). We refer to \cite{ADM22} for a higher-dimensional extension of the elliptic GinUE based on the harmonic oscillator Hamiltonian in general complex dimensions.

As our third and final application under this heading, denote by $\tilde{G}$ a rectangular $n \times N$ ($n \ge N)$ GinUE matrix, and by $U$ and $N \times N$ Haar distributed \alert{unitary} random matrix. From these form $X = (G^\dagger G)^{1/2} U$. Matrices of this type are said to form the induced GinUE
\cite[\S 2.4]{BF25}. The corresponding eigenvalue \alert{PDF} is proportional to (\ref{E.5}), now multiplied by the extra factor $\prod_{l=1}^N |z_l|^{2(n-N)}$, and the corresponding normalisation is
$C_{n,N} = N! \pi^N \prod_{j=1}^N (n-N+j-1)!$ \cite{FBKS12}. 
Writing $n = (1+\alpha)N$, $\alpha > 0$ we see that the potential in this new PDF is, up to an additive constant, that given in (\ref{3.4ai}) with $R^2 = (1 + \alpha)N$, $c^2 = \alpha/(1 + \alpha)$. From the analogy, these latter values imply that the support of the eigenvalue density is to leading order the annulus $\sqrt{\alpha} < |z| < \sqrt{1 + \alpha}$, and that in this region it has value $1/\pi$. Also, making use of (\ref{3.4bi}), for the induced GinUE configuration integral the analogy predicts
$$
\log {1 \over N!} Z_N \sim  {N^2 
\over 2 } (1 + \alpha) \log ( (1 + \alpha) N) - {1 \over 4} N^2 (3 + 4 \alpha)  + {\rm O}(N),
$$
which indeed is consistent with the asymptotics following from the explicit form of $C_{n,N}$.

\section{Electrostatic Potentials Due to Surface Charges}\label{S4}

\alert{A fundamental setting in electrostatics is the distribution of charge on a conducting surface --- often referred to as the equilibrium charge distribution. 
A characterising feature is that the potential on the surface is a constant. Of interest is the explicit form of the surface density, and the corresponding potential. This is calculated explicitly for the sphere in dimension $d$ and the surface of the hyperellipse. It is shown how a projection method can be used to deduce the equilibrium measure for the Coulomb potential on a $d-1$ dimensional ball. In the two-dimensional case more can be said using conformal mappings and Green function theory. A viewpoint of the equilibrium measure, combined with a scaling property of Boltzmann factor for the logarithmic potential, allows for the computation of the first three terms of the asymptotic expansion of the free energy of the log-gas confined to a contour. This is similarly true of the corresponding version of the two-dimensional one-component plasma, and extends to the consideration of the effect of an impurity charge which in turn relates to Fisher-Hartwig asymptotics.
Conformal mappings and Green functions arise when considering fluctuation formulas for linear statistics in this setting, and also for infinite density limit of the two-dimensional one-component plasma.
}

\subsection{Spherical and Ellipsoid Surfaces}
In the setting of the  equilibrium charge distribution on the surface of $\Omega$, one does not take the viewpoint of first imposing a uniform background on this surface. Doing so is a different setting, albeit one which occurs in applications; recall the first paragraph of \S \ref{S1.2}.

The simplest case of the  equilibrium charge distribution
is that of a spherical conducting surface $\partial B_R$ in $\mathbb R^d$. Let the total charge on the surface be $Q$, \alert{and let} $c_d$ denote the surface area of the unit sphere in $\mathbb R^d$ as used in (\ref{1.1}). The surface density $\sigma_d(\vec{r})$ say, which is supported on $\partial B_R$, must by symmetry be equal to the constant $Q/(R^{d-1} c_d)$,
\alert{determined by the requirement} that it integrates to give $Q$. The corresponding potential is then
\begin{equation}\label{4.1}
V(\vec{r}) = {Q \over R^{d-1} c_d} \int_{\partial B_R}
\Phi_{\rm d}(\vec{r},
\vec{r} \, ') \, dS_{\vec{r}\,'},
\end{equation}
where $dS_{\vec{r}\,'}$ denotes the volume element associated with $\partial B_R$ in the coordinate $\vec{r}\,'$.
As in Proposition \ref{P3.1} the symmetry of the sphere tells us that $V(\vec{r}) = V(r)$
(i.e.~is a function only of the length of $\vec{r}$), however in distinction to the circumstance of that result, we cannot proceed to compute (\ref{4.1}) using the Poisson equation characterisation of the Coulomb potential due to the incompatibility with the resulting delta function and the surface integral. An alternative way to proceed is to regard the surface of the sphere as a limit of the spherical annulus geometry, \alert{as} considered in the paragraph below the proof of Proposition \ref{P3.1}. This shows
\begin{equation}\label{4.2}
V(\vec{r}) =
\begin{cases} Q \Phi_{\rm d}(R), & \vec{r} \in B_R, 
\\
Q \Phi_{\rm d}(r), & {\rm otherwise},
\end{cases}
\end{equation}
in accordance with Newton's shell theorem. \alert{H}ere one has to appropriately account for the limit of the thickness of the annulus going to zero; see the working below \alert{in the case of the} more general hyperellipsoid.

\alert{Indeed we turn our attention now to the case of the hyperellipsoid}.
It is shown in \cite{C+20} that \alert{determinantion of} the surface charge  also yields to an analysis based on the annulus like region $\Omega \backslash c \Omega$ in the limit $c \to 1^-$ with $\Omega$ denoting the interior of the hyperellipsoid.

\begin{proposition}\label{P4.1}
In the above setting, where the hyperellipsoid surface is given by \eqref{3.5}, we have
\begin{equation}\label{4.3}
\sigma_d(\vec{r}) =
{Q \over c_d a_1 \cdots a_d}
{1 \over (\sum_{j=1}^d x_j^2/a_j^4)^{1/2}},
\end{equation}
supported on $\partial \Omega$. Furthermore, in terms of the quantities (\ref{3.12}), 
$V(\vec{r})$ for $\vec{r}$ outside of $\partial \Omega$ is given by 
\begin{equation}\label{4.4} 
 V(\vec{r}) =
  Q(d-2)
 \int_0^\infty   {1
 \over \sqrt{S_0(\lambda})}
 \mathbbm 1_{S_1(\lambda) < 1} \,  d \lambda,
  \end{equation}
  valid for $d > 2$; cf.~(\ref{3.13}).
  Inside of $\partial \Omega$, $V(\vec{r})$ is a constant.
  \end{proposition}

  \begin{proof}
  The fact that $V(\vec{r})$ is a constant in the interior follows from 
  the discussion of the paragraph including 
  (\ref{3.13a}).
In relation to the result for $\sigma_d(\vec{r})$,
  the crucial formula is that for a surface $F(\vec{r}) = 0$, \alert{one has for  the Dirac delta function the identity}
 \begin{equation}\label{4.5} 
 \delta(F(\vec{r})) = {1 \over | \hat{n} \cdot \nabla F(\vec{r} |}
 \delta(u - u_0),
  \end{equation}
  where $u$ is the direction normal to the surface, and $u=u_0$ when $F(\vec{r})=0$ with outward unit normal vector $\hat{n}$.
  Beginning with the distribution in $\mathbb R^d$, $$(q/c_d a_1\cdots a_d)
  \delta (F(\vec{r})),$$
 with $F(\vec{r}) = 0$
  the equation determining the ellipsoid, use of (\ref{4.5}) and integration over $u$ gives (\ref{4.3}).
  A calculation analogous
   to that of the proof of Proposition \ref{P3.2}, now using the Fourier integral form of the delta function instead of (\ref{3.13}), establishes (\ref{4.4}). 
  \end{proof}

  \subsection{Projections of Spherical and Ellipsoidal Uniformly Charged Bodies}
As an introductory example, consider the ball $B_R$ in dimension $d=2$, with its boundary (the circle of radius $R$) uniformly charged with total charge $-N$ --- hence the density of this uniform charge is $\rho_{\rm b} = - N/2 \pi R$. We know that inside $B_R$ the potential in this setting is given by 
\alert{the first case of} (\ref{4.2}). Suppose now that we project each infinitesimal uniform charge surface charge density $\rho_{\rm b} d \theta $ to the real axis by setting
$x = R \cos \theta$. \alert{This gives} the  one-dimensional distribution
according to the change of variables
\begin{equation}\label{S.1}
\mu(x) dx := 2 \rho_{\rm b} 
 d \theta = 2 \rho_{\rm b}  {dx \over R (1 - (x/R)^2)^{1/2}} \mathbbm 1_{|x| < R}
\end{equation}
(the factor of 2 is due to the two half circles being projected to the same point on the real axis).

The projected charge density (\ref{S.1}) is no longer uniform. Instead we recognise its functional form as that resulting from the Jacobi ensemble in random matrix theory
\cite[\S 3.6]{Fo10}, specified by replacing the quadratic potential in (\ref{1.8}) by $-(a/2) \log (1 - \lambda) -(b/2) \log(1 + \lambda)$, supported on $|\lambda| < 1$. This being independent of $a,b$ we can set the parameters to zero to conclude that $\mu(x)$ is the equilibrium measure with respect to the logarithmic potential on the interval $(-1,1)$,
in the sense that the equation (\ref{1.10})
holds with the quadratic potential deleted (and $c=1$).
 This is not a coincidence, with it being possible to  
 deduce the equilibrium measure of  the Coulomb potential on the $d-1$ dimensional ball by projection for general $d \ge 2$.

We give the details for $d=2$.
The argument, which in the web resource \cite{Mc19} is recalled and traced back to W.~Thompson (Lord Kelvin), and can be found too in the further web resource \cite{FoX}, proceeds by considering
the electric field generated at a point $P$ on the real axis due to two ``opposite''
  small arcs on the surface,
 specified by them being connected by two straight lines joining opposite ends of the two arcs, both passing through $P$. 
 \alert{Following} the geometric argument given by Newton in relation to his shell theorem, since the solid angles are the same, the electric field at $P$ due to these two arcs is equal and opposite, and so cancels.

Crucially, from the geometry of the 
construction in the limit that the sizes of the arcs go to zero,
the ratio of distances from the two arcs to $P$ remains unchanged upon replacing $P$ by its projection to the $x$-axis. This gives that the total electric field (or equivalently potential) at $P$ can be calculated by projecting the charge density from the boundary of the disk to the real line,
giving the new equilibrium surface charge density. This same argument applies to the projection of the uniformly charged surface of a ball in dimension $d$, implying the equilibrium charge distribution $\sigma_{d-1}(\vec{r})$ on a ball in dimension $d-1$ with respect to the Coulomb potential in dimension $d$.

\begin{proposition}
Let $\vec{r}$ be a point inside the ball $B_R$ of dimension $d-1$, let $\sigma_{d-1}(\vec{r} \,') = - (2 \rho_{\rm b}/R) (1 - (|\vec{r} \, '|^2/R)^2)^{-1/2} \mathbbm  1_{|\vec{r} \, '| < R}$, and let $\Phi(\vec{r}, \vec{r} \, ')$ denote the
Coulomb potential in $\mathbb R^d$. We have
\begin{equation}\label{S.2}
 \int_{B_R |_{d \mapsto d - 1}} \sigma_{d-1}(\vec{r} \, ') \Phi(\vec{r}, \vec{r} \, ') \, d \vec{r} \,' = C,
\end{equation}
where $C$ is a constant.
\end{proposition}

The projection method applies equally as well to the hyperellipsoid surface (\ref{3.5}), with it sharing the property with the ball $B_R$ of its equilibrium surface distribution creating a constant potential throughout the interior;
see \cite{Mc19}, \cite{C+20} for more on the projection operation in this setting.

\begin{remark} ${}$ \\
1.~There are other potential problems \alert{(albeit not equilibrium measures)} which have as their solution the charge density 
$\rho(\vec{r})$ the functional form
proportional to $(1 - (|\vec{r} \, '|^2/R)^2)^{-1/2}$ in dimensions $d \ge 2$. Let $\Psi_s(\vec{r},\vec{r}\, ')$ be the Riesz potential (\ref{1.11}). Then we have from \cite[Th.~4]{CSW22} that
\begin{equation}\label{S.3}
\int_{B_R} \rho(\vec{r}) 
\Psi_{d-3}(\vec{r}, \vec{r} \,') \, d \vec{r} \, ' = - \gamma |\vec{r} |^2, \quad \vec{r} \in B_R
\end{equation}
for some $\gamma > 0$ related to $R$.
\\
2.~Consider the case of a uniformly charged hyperellipsoid from Proposition \ref{P3.2}, which we denote $\Omega_d$. 
We know that the potential \alert{in the interior} is a quadratic; recall (\ref{3.13a}). Suppose now the parameter $a_d$ in the equation (\ref{3.5}) determining this domain is small. To leading order the contribution to the 
$d$-dimensional integral determining the potential 
$-\rho_{\rm b} \int_\Omega\Phi_{d}(\vec{r}, \vec{r} \,') \, d\vec{r} \, '$,
with $\vec{r}$ restricted to $x_d = 0$ (identified as in $\mathbb R^{d-1}$, \alert{and then to be denoted} by $\vec{s}$), can be simplified by carrying out the integration over $x_d'$. This gives \alert{gives rise to} a factor of
$2 a_d \mu(\vec{s} \, ')$, $\mu(\vec{s} \, ') = (1 - (x_1^2/a_1^2 + \cdots + x_{d-1}^2/a_{d-1}^2))^{1/2}$  in the integrand,  \alert{it} replaces the factor
$\Phi_{d}(\vec{r}, \vec{r} \,')$ by
$\Phi_{d}(\vec{s}, \vec{s} \,')$, and the integration domain is now $\Omega_{d-1}$. Recalling too 
that $\rho_{\rm b} = N/|\Omega_{\rm d}|$,
we see that this limiting procedure  gives from (\ref{3.13a}) the \alert{hyperellipsoid domain, charge profile $\mu(\vec{s})$}, potential evaluation in dimension $d-1$,
\begin{equation}\label{4.9x}
\sum_{j=1}^{d-1} \alpha_j x_j^2 + \alpha_0 =
-{2 N \Gamma(d/2+1) \over \pi^{d/2} a_1 \cdots a_{d-1}} 
\int_{\Omega_{d-1}}
\Phi_d(\vec{s},\vec{s}\, ') \mu(\vec{s} \, ')
\, d \vec{s} \,'.
\end{equation}
This result for $d=3$ can be traced back to the classic work of Hertz on contact mechanics (see 
the recent review on the analogy between contact mechanics and electrostatics \cite[\S 2.3.2]{BVC23} for a citation) and  
\cite[\S 5.1]{Ma13} for a precise statement.

In the special case of a ball and for general $d \ge 2$ (\ref{4.9x}) can be recognised as 
 a special case of more general identities \cite[Lemma 4.1]{BIK15},
\cite[Cor.~4 and Remark 1]{DKK17}, \cite[Lemma 2.4]{GCO23}, \cite[Th.~1.4]{BFMS25}.
For $d=2$, 
we can read off the value of $\alpha_1, \alpha_0$ from (\ref{3.13c}) to deduce (after a change in notation)
\begin{equation}\label{4.9y}
{x^2 \over 2} + {a^2 \over 2} \log {a \over 2} - {a^2 \over 2} =
{a \over \pi}
\int_{-a}^a \log |x - s| \, (1 - s^2/a^2)^{1/2} \,
ds, \quad |x| < a,
\end{equation}
which is consistent with the Wigner semi-circle solution of equation
(\ref{1.10}); see \cite[Prop.~1.4.3]{Fo10}. That after a projection onto the real axis a constant density on the ellipse implies a semi-circle law on the real line was first observed in \cite{SCSS88}.

\end{remark}

  \subsection{The Logarithmic Potential and Charge Density on Conducting Surfaces in the Plane}\label{S4.2}

  Let $\partial \Omega$ denote the boundary of a compact set on $\mathbb R^2$, which we take as being conducting. Distribute a total charge $Q$ on $\partial \Omega$ according to the \alert{normalised} surface charge density $\sigma(\vec{r})$ so that it is an equipotential with respect to the logarithmic potential
\begin{equation}\label{U1}
V(\vec{r}) = - Q \int_{\partial \Omega}
\log |\vec{r} - \vec{r}'|
\sigma(\vec{r}\, ') \, ds,
\end{equation}
where $ds$ relates to the length element on $\partial \Omega$. Taking a viewpoint in the complex plane we can write
\begin{equation}\label{U2}
V(z) = - Q \int_{\partial \Omega}
\log |z - z'|
\sigma(z \,') \, |dz'|.
\end{equation}
Characterising properties of $V(z)$ are that it satisfies the Laplace equation for $z \in \mathbb C \backslash \bar{\Omega}$, behaves as $-Q\log|z|$ at infinity, and is equal to a constant value, $Q V_{\partial \Omega}$ say, for $z \in \partial \Omega$. 

There is further significance to both $\sigma(z)$ and $V_{\partial \Omega}$ in this setting. Thus introducing the logarithmic energy
$-  \int_{\partial \Omega} |dz| \, \mu(z)
\int_{\partial \Omega} |dz'| \, \mu(z')
\log |z - z'|$, for general measures $\mu(z)$ supported on $\partial \Omega$, one has that $\mu(z) = \sigma(z)$ is such that this quantity is minimised; see \cite{ST97}. Using the notation introduced below (\ref{U2}), the minimum value of the logarithmic energy is $V_{\partial \Omega}$. This in turn specifies the logarithmic capacity ${\rm cap} \, (\partial \Omega)$ according to
$ V_{\partial \Omega} = - \log 
{\rm cap} \, (\partial \Omega).
$

Analogous to the discussion relating to
(\ref{3.21}),
the formulation (\ref{U2}) allows complex variables techniques to be used to specify $V(z)$ for several geometries.
Important here is the conformal map $w = \zeta(z)$ from 
$\mathbb C \backslash \bar{\Omega}$ to the exterior of the unit disk in the complex $w$-plane. This can be chosen to have the form at infinity
\begin{equation}\label{U3}
\zeta(z) = {z \over c} + a_0 + {a_1 \over z} + \cdots, \qquad c > 0.
\end{equation}
We then set $V(z) - Q V_{\partial \Omega} = - 
Q \log |\zeta(z)|$. We see that $- \log |\zeta(z)|$ must satisfy Laplace's equation for $z \in \mathbb C \backslash \bar{\Omega}$,
must equal zero for $z \in \partial \Omega$, and must behave as $\log|z|$ at infinity (these last two conditions tell us that in $(\ref{U3})$,
$\log c = - V_{\partial \Omega}$). These properties in fact uniquely characterise
$- \log |\zeta(z)|$
as the Green function $g_{\partial \Omega}(z,\infty)$ associated with $\bar{\Omega}$ \cite{ST97}.

To see the use of this we typically start with the inverse mapping $z = \xi(w)$, chosen so that $|w|=1$ gives  $\partial \Omega$. The standard example is the limit of a slim ellipse, when it degenerates to the unit interval $[-1,1]$ on the real axis. Then the particular Joukowski map $z = {1 \over 2} (w + 1/w)$ has the sought property. Solving for $w$ in terms of $z$ and choosing the root compatible with (\ref{U3}) gives
$w = \zeta(z) = z + \sqrt{z^2 - 1}$. We conclude that in this setting
\begin{equation}\label{U4}
g_{\partial \Omega}(z,\infty) = - \log | z + \sqrt{z^2 - 1}|.
\end{equation}

In general the surface charge density $\sigma(z)$ relates to the conformal map $\zeta(z)$ by $\sigma(z) = {1 \over 2 \pi} |\zeta'(z)|$; see e.g.~\cite[\S 3]{Sa10}. One way to see this is to make use of what is known as Green's third identity specific to harmonic functions in the complex plane, 
$$
g_{\partial \Omega}(\zeta) - V_{\partial E} = -{1 \over 2 \pi} \int_{\partial \Omega}
\Big (\log |z - \zeta | \Big ) {\partial g(z) \over \partial n} \, |dz|,
$$
where in computing the derivative in the normal direction to the contour, the fact that $g(\zeta)$ is a constant along the contour is to be used.
An alternative approach is to first take the (complex) derivative with respect to $z$ in the real part of (\ref{U2}), which removes the absolute value sign from the logarithm, then calculating difference of the limit $z \to z_0 \in \partial \Omega$ from the outside, and from the inside, as required by  the Sokhotski-Plemelj formula. Applying this theory gives that $\sigma(x)$ in relation to (\ref{U4}) is given by ${1 \over \pi} (1-x^2)^{-1/2}\mathbbm 1_{|x|<1}$.

A more general Joukowski map can be used to treat the case that $\Omega$ is an ellipse.

\begin{proposition}\label{P4.2}
Set $z = {R \over 2}(w + {c^2 \over w})$, $R > 0$, $0 \le c \le 1$, so that the exterior of the ellipse in the $z$-plane
with semi-axis lengths $a = {R \over 2} (1 + c^2)$, $b = {R \over 2} (1 - c^2)$, is mapped to the exterior of the unit circle in the $w$-plane. We have
\begin{align}
\begin{split}
\label{U5}
& g_{\partial \Omega}(z) = - 
\log |(z/R) + \sqrt{(z/R)^2 - c^2} |,
\\
& \sigma(z) = {1 \over \pi  } \bigg | {1 \over (z^2 - (Rc)^2)^{1/2} } \bigg |, \quad V_{\partial \Omega} = \log {R \over 2}, 
\end{split}
\end{align} 
where in the first of these formulas we require $z$ to be outside of the ellipse, while in the second we require $z$ to be on the boundary.
\end{proposition}

We note that a functional form of $\sigma(z)$ has already been given in
(\ref{4.3}) of Proposition \ref{P4.1}
(set $Q=1$, $d=2$), which after appropriate parameterisation can be checked to agree.

\subsection{The Green Function $g_{\mathbb C \backslash \bar{\Omega}}(z,w)$}
The viewpoint for our applications \alert{below} is that $\Omega$ is a conducting domain (which may be infinite) with surface that is held at zero potential. A unit charge is placed at
$w \in \mathbb C \backslash \bar{\Omega}$, and we denote by
$g_{\mathbb C \backslash \bar{\Omega}}(z,w)$ the electrostatic potential at $z$ due to this charge and the effect \alert{on} the conducting medium. More precisely, 
$g_{\mathbb C \backslash \bar{\Omega}}(z,w)$ is the real valued function satisfying the two-dimensional Poisson equation
\begin{equation*}\nabla_z^2
g_{\mathbb C \backslash \bar{\Omega}}(z,w) = - 2 \pi \delta(z-w),
\end{equation*}
outside of the conducting medium,
subject to the boundary condition that it vanishes on $\partial \Omega$. To make the Green function unique it is required that it be symmetric in $z,w$.

Well known and easy to verify is that for $\Omega = B_R$ (the disk of radius $R$),
\begin{equation}\label{U6}
g_{\mathbb C \backslash \bar{B}_R}(z,w) =
- \log {|z - w| \over |1 - z \bar{w} /R^2|}.
\end{equation}
Rearranging the denominator
$\log |1 - z \bar{w} /R^2| =
\log |z - R^2/\bar{w}| + \log |\bar{w}/R^2|$, one sees that the effect of the \alert{conducting} disk to create an image charge of opposite sign at the inversion point $R^2/\bar{w}$ within the disk.
The conformal map $\zeta(z) = (z+i)/(z-i)$ maps to upper half plane $\mathbb H$ to the exterior of the unit disk. Substituting in (\ref{U6}) with $R=1$ gives the further well known result
\begin{equation}\label{U7}
g_{\mathbb H}(z,w) =
- \log {|z - w| \over |z - \bar{w}|},
\end{equation}
which is easy to independently deduce using the method of images (i.e.~the effect of the \alert{conducting} half plane is to create a negative charge at $\bar{w}$ in the half plane). Generally, if the conformal map from the outside of $\Omega$ to the outside of the unit circle is $\zeta(z)$, it follows from (\ref{U6}) that
\begin{equation}\label{U8}
g_{\mathbb C \backslash \bar{\Omega}}(z,w) =
- \log {|\zeta(z) - \zeta(w)| \over |1 - \zeta(z) \overline{\zeta(w)}|}.
\end{equation}
We refer the reader to \cite{CM07} and \cite[\S~2]{ABK24} for further discussions on Green’s functions in various domains, particularly those beyond the simply connected case, where Jacobi theta functions and their extensions---such as the Schottky--Klein prime functions or Riemann theta functions---play a natural role in their representation.

\begin{remark}
Suppose the conducting medium is $B_R$ in $d=3$. Let $g_{\mathbb R^3 \backslash \bar{B}_R }
(\vec{r},\vec{r} \, ')$ denote the Green function at $\vec{r}$
due to a point charge  at
$\vec{r} \, '$ and the response in the conducting medium. While complex variable methods are no longer applicable, the method of images can be used to show
\begin{equation}\label{U9}
g_{\mathbb R^3 \backslash
\bar{B}_R}(\vec{r}, \vec{r}\,') = {1 \over |\vec{r} - \vec{r} \, '|} -
{R \over |\vec{r} \, '|
|\vec{r} - R^2 \vec{r}\, ' / |\vec{r}\, '|^2 |}.
\end{equation}
As another well known example in $\mathbb R^3$ for which the Green function can be computed using the method of images, let $\Omega = \mathbb R^3_-= \{ (x,y,z) \in \mathbb R^3: \, z < 0 \}$. For this  conducting region we have
\begin{equation}\label{U10}
g_{\mathbb R^3 \backslash
\bar{\Omega}}(\vec{r}, \vec{r}\,') = {1 \over |\vec{r} - \vec{r} \, '|} -
{1 \over |\vec{r} - \vec{r}_-^{\,\prime}| },
\end{equation}
where $\vec{r}_-^{\,\prime} = (x',y',-z')$. Such results are covered in standard texts such as \cite{Ja75}.
\end{remark}

\subsection{Applications}

\subsubsection{Log-Potential Coulomb Gas on a Closed Contour --- Free Energy Expansion}

Let $\partial \Omega$ be a closed contour, corresponding to the outer boundary of a compact set $\Omega$ in $\mathbb R^2$ (or in $\mathbb C$).
For $N$ unit charges repelling via the logarithmic potential and contained in the closure of $\Omega$, it is a known theorem  that the minimum energy configuration has each particle positioned on $\partial \Omega$ --- this follows from the maximum modulus principle for analytic functions; see e.g.~\cite{Sa10}). In the large $N$ limit the resulting normalised surface charge density must then be an equipotential, and thus is identified as $\sigma(\vec{r})$ from
\S \ref{S4.2}. An alternative way to arrive at $\sigma(\vec{r})$ starting with a statistical mechanical system is to confine the log-potential Coulomb gas \alert{without a background} to $\partial \Omega$, then to take the large $N$ limit. In either case the circumstance is distinct from that of the one-component plasma as there is no background. Moreover we are interested in the large $N$ limit with $\partial \Omega$ fixed, which is distinct from the thermodynamic limit when both $N$ and the volume go to infinity simultaneously with the density constant. 

For purposes of predicting the leading large $N$ asymptotics, we propose to identify an equivalent system which does have a background and which furthermore permits a scaling allowing for the thermodynamic limit. This equivalent system starts by imposing a background surface charge density $-N \sigma(\vec{r})$ on $\partial \Omega$. Since by the characterisation of $\sigma(\vec{r})$ as implying that $\partial \Omega$ is a equipotential, this does not alter the normalised charge density of mobile particles equaling $\sigma(\vec{r})$. With $V_{\partial \Omega}$ defined as below (\ref{U2}), this implies the contribution to the Boltzmann factor from the background-background, and background particle interactions
$e^{\beta N^2 V_{\partial \Omega}/2}$. We furthermore scale $\partial \Omega \mapsto N \partial \Omega$ so as to be considering the one-component log-gas plasma confined  to this latter contour with Boltzmann factor
$e^{\beta N^2 V_{\partial \Omega}/2}
\prod_{1 \le j < k \le N} |z - z '|$. For this we expect the corresponding free energy to be extensive. After rescaling the contour back to the original by changing variables in the particle coordinates, we are lead to the predicted large $N$ asymptotic formula
\begin{align}
\begin{split}
\label{U11}
&\quad \log {1 \over N!} \int_{\partial \Omega} |d z_1| \cdots \int_{\partial \Omega} |d z_N| \,
\prod_{1 \le j < k \le N}| z_k - z_j|^\beta 
\\
& \sim - {\beta N^2 \over 2} \log N 
-{\beta N^2 \over 2}  V_{\partial \Omega} + \Big ( {\beta \over 2} - 1 \Big ) N \log N,
\end{split}
\end{align} 
up to terms of order $N$. This indeed agrees with a known rigorous result \cite{Jo88}, \cite{CJ25} (these references extend the asymptotic expansion to give too the order $N$ and the constant term --- see also \cite{WZ22}).
One remarks that the derivation of (\ref{U11}) applies equally as well in the circumstance that the closed curve $\partial \Omega$ is replaced by an open contour $\gamma$ say. The validity of the predicted formula is confirmed by the recent rigorous work \cite{CJV25}, which moreover extends the asymptotic expansion through to the constant term; see also \cite{GKL24} and references therein for related works.

\begin{remark}
    \alert{In deducing (\ref{U11}) essential use has been made of the scaling property of the Boltzmann factor in the logarithmic case with respect to dilation. This is not the case for the Coulomb potential in higher dimensions, for which our argument therefore breaks down.}
\end{remark}

\subsubsection{Log-Potential Coulomb Gas on a Closed Contour --- Fluctuation Formulas}

\alert{Consider} the case of $\partial \Omega = \partial B_1$ which we interpret as the unit circle in the complex plane. \alert{T}he Boltzmann factor of the corresponding log-gas model with $\beta = 2$, $\prod_{1 \le j < k \le N}| e^{i \theta_k} - e^{i \theta_j} |^2$, is proportional to the joint eigenvalue PDF for Haar distributed unitary random matrices; see e.g.~\cite{DF17}. This is a determinantal point process \cite[Ch.~5]{Fo10} and so has its correlation functions determined by a single kernel function
\begin{equation}\label{U12}
K_N(\theta,\theta') = {1 \over 2 \pi}
{\sin N (\theta - \theta')/2 \over 
\sin  (\theta - \theta')/2}.
\end{equation}
Specifically, we have for the so-called truncated two-point correlation function (see \cite[\S 5.1.1]{Fo10}) $\rho_{(2),N}^T(\theta,\theta') = - |K_N(\theta,\theta')|^2
$. Generally the truncated two-point correlation is fundamental to fluctuation formulas for so-called linear statistics \alert{as seen in the formula} \cite[Prop.~2.6]{Fo23}
\begin{align}
\begin{split}
\label{U13}
&\quad {\rm Cov} \, \Big ( \sum_{l=1}^N f(\theta_l),\sum_{l=1}^N g(\theta_l) \Big )  
\\
& = - {1 \over 2}  \int_0^{2 \pi} d\theta   \int_0^{2 \pi}  d\theta' \, (f(\theta) - f(\theta')) (g(\theta) - g(\theta'))   \rho_{(2),N}^T(\theta,\theta').  
\end{split}
\end{align} 
Squaring (\ref{U12}), noting that
$(\sin N (\theta - \theta')/2)^2 = {1 \over 2} (1 - \cos N (\theta - \theta'))$ and substituting as appropriate, we observe the limiting form \alert{for the present log-gas model} (see e.g.~\cite[Eq.(2.23)]{Fo23})
\begin{align}
\begin{split}
\label{U14}
&\quad \lim_{N \to \infty} 
{\rm Cov} \, \Big ( \sum_{l=1}^N f(\theta_l),\sum_{l=1}^N g(\theta_l) \Big ) 
\\
& =  {1 \over (4 \pi)^2}  \int_0^{2 \pi} d\theta   \int_0^{2 \pi}  d\theta' \, {(f(\theta) - f(\theta') 
(g(\theta) - f(\theta')
\over 
 (\sin (\theta - \theta')/2 )^2},
\end{split}
\end{align} 
due to the oscillatory term causing cancellations in the integral to leading order and so not contributing.

It is instructive to compare (\ref{U14}) to the analogous result in relation to the $\beta = 2$ log-gas model with Boltzmann factor $\prod_{1 \le j < k \le N}| z_k - z_j |^2$, \alert{the domain now $B_1$ rather than $\partial B_1$}. Up to normalisation, the latter arises in random matrix theory as the eigenvalue PDF of the top $N \times N$ sub-block of an $(N+1) \times (N+1)$ Haar distributed random matrix \cite{ZS99}. This is again a determinantal point process, now with kernel
\begin{equation}\label{U15}
K_N(z,z') = 
 {1 \over \pi} \sum_{j=1}^{N} j (z \bar{z} ')^{j-1} = {1 \over \pi}
 {d \over d w} {1 - w^{N+1} \over
 1 - w} \bigg |_{w = z \bar{z} '}
 \mathop{\sim} \limits_{N \to \infty} - {N \over \pi} {w^N \over 1 - w} \bigg |_{w = z \bar{z} '},
 \end{equation}
where the validity of the large $N$ form requires $w \ne 1$.
From this functional form, it follows from integration by parts that
\begin{equation}\label{U16}
\int_0^1 r dr  \int_0^1 r' dr' \,
| K_N(r e^{i \theta}, r' e^{i \theta'}) |^2 \sim \Big ( {1 \over 2 \pi} \Big )^2 {1 \over |1 - e^{i (\theta - \theta')}|^2},
 \end{equation}
which we recognise as $1/2$ of the smoothed part of $\rho_{(2),N}^T(\theta,\theta')$ as implied by (\ref{U12})
(by the smoothed part, we mean after application of the identity noted below (\ref{U13}) with the oscillatory term discarded). In particular, this provides another way to arrive at the limiting boundary fluctuation formula (\ref{U14}); albeit multiplied by $1/2$
(see the subsection below for more on this modification).

There is a link between (\ref{U14}) and $g_{\mathbb C \backslash B_1}(z,w)$,
\begin{align}
\begin{split}
\label{U17}
&\quad g_{\mathbb C \backslash B_1}(z,w) +
\log |z - w|
\\ 
&= \beta \bigg (  
\lim_{N \to \infty} 
{\rm Cov} \, \Big ( - \sum_{l=1}^N \log | e^{i \theta_l} - z|,- \sum_{l=1}^N \log | e^{i \theta_l} - w| \Big ) + \log  |zw| \bigg ),
\end{split}
\end{align}
for $|z|,  |w| > 1.$
This follows from a linear response argument, with the charge at $z$ regarded as having infinitesimal strength $\delta q$,
and considering the change in the linear statistic
$-\sum_{l=1}^N \log |w-z_j|$. From an electrostatics viewpoint, in the infinite density limit this should be the macroscopic potential at $w$ induced on the conductor by the charge at $z$. On the other hand, from a statistical mechanics viewpoint, one first includes in the total energy a factor $-\delta q \sum_{l=1}^N \log |z-z_j|$ to account for the test charge of strength $\delta q$ at $z$, then expands to Boltzmann factor to leading order in $\delta q$ to arrive at the RHS (the factor of
$\delta q$ cancels out from both sides); see \cite{Ja95}, \cite[\S 14.3]{Fo10}.
For a direct verification, one requires knowledge that the RHS of (\ref{U14}) can be written in the form $ \sum_{n = - \infty}^\infty |n| f_n g_{-n}$, $ f_n := {1 \over 2 \pi }\int_0^{2 \pi} f(\theta) e^{-i n\theta} \, d \theta$ (and similarly $g_n$); see e.g.~\cite[Eq.~(2.14)]{Fo23}.

\alert{The present viewpoint tell us}
the identity (\ref{U17}) must remain true for $B_1$ replaced by general compact $\Omega$, $|z|, |w| > 1$ replaced by $z,w \in \mathbb C \backslash \bar{\Omega}$, and each $e^{i \theta_l}$ by a point on
$\partial \Omega$.
From (\ref{U13}), with the integrations now over $\partial \Omega$, it follows that
$g_{\mathbb C \backslash \Omega}(z,w)$ --- which is given explicitly by (\ref{U8}) --- must be related to the smoothed, limiting form of $|K_N(z,w)|^2$, which we will denote $|K_\infty(z,w)|^2$. Indeed, results from \cite[Cor.~1.4]{AC23}, in the setting that the particles form a droplet with support $\Omega$ and which relate the analogue of $K_\infty(z,w)$ to the so-called exterior Szeg\H{o} kernel,  predict  
\begin{equation}\label{U18}
|K_\infty(z,w)|^2 =
{1 \over 2 \pi^2}
{|\zeta'(z) \zeta'(w)| \over|1 - \zeta(z) \overline{\zeta(w)} |^2}, \quad z,w \in \partial \Omega.
\end{equation}
We refer to \cite[Eq.~(1.24)]{BY23} for an analogous formula in the case where the droplet is not connected.
With $z = \xi(u)$ denoting the inverse map of $u = \zeta(z)$, substituting (\ref{U18}) as appropriate in (\ref{U13}) and changing variables gives for the predicted limiting fluctuation formula
\begin{align}
\begin{split}
\label{U19}
&\quad \lim_{N \to \infty} 
{\rm Cov} \, \Big ( \sum_{l=1}^N f(z_l),
\sum_{l=1}^N g(\bar{w}_l)
\Big ) 
\\
&=  {1 \over 4 \pi^2} 
\int_{|u|=1} | du|
\int_{|v|=1} | dv| \,
 {(f(\xi(u)) - f(\xi(v))
 (g(\overline{\xi(u)}) - g(\overline{\xi(v)})
 \over |u - v|^2} ;
\end{split}
\end{align} 
cf.~\cite[Remark 1.5]{ADM24}. However, it is not immediate to verify from this the requirement
(\ref{U17}) generalised to
the case of 
$g_{\mathbb C \backslash \bar{\Omega}}(z,w)$.
We remark too that different formulas for (\ref{U19}) have been given in 
\cite{Jo88}, \cite{WZ22} and \cite{Jo22}. 

The above considerations for the log-gas on a contour at $\beta = 2$ can be extended in two ways. One is to allow for a general inverse temperature $\beta > 0$. This is simple to account for, 
as $\beta$ only appears as a factor on the RHS of (\ref{U17}). To compensate, the RHS of (\ref{U14}) and (\ref{U19}) are to be multiplied by $2/\beta$. Doing the same to the RHS of (\ref{U18}) gives the predicted functional form of the (negative of the) limiting smoothed truncated two-point correlation.

The other extension is to consider the circumstance of an interval rather than closed contour. For the interval $[-1,1]$ we know from the text above (\ref{U4}) that the 
\alert{conformal map} is such that $\xi(e^{i \theta}) = \cos \theta$. Substituting in (\ref{U19}) gives a known formula for the covariance of a linear statistic in this setting
\cite[Eq.~(4.15)]{La18}, although somewhat surprisingly, with an extra factor of two. 
This does not contradict the conjectured validity of (\ref{U19}) for a closed contour due to the limiting  case of an interval is no longer in this class.
Furthermore, in the next subsection we will be lead back to 
(\ref{U19}) with the prefactor multiplied by $1/2$, when we consider the region $\Omega$ as containing a neutralising background. We remark that the recent work \cite{CJV25} considers the circumstance of the log-gas confined to a Jordan arc, thus generalising the case of an interval. Other recent developments include linear statistics for high dimensional Coulomb gases and one-dimensional Riesz gases \cite{DLMS24,LS25}. 

\subsubsection{The 2d Log-Potential Coulomb Gas with a Background}
Here we are imposing a uniform background in a compact region $\Omega$, charge density $-\rho_{\rm b} = - N/|\Omega|$. In keeping with the situation of the previous subsection, we want to keep $\Omega$ fixed and consider the limit $N \to \infty$. By doing so allows for the application of methods of electrostatics for the study of the asymptotics of the free energy, the surface correlations and associated fluctuation formulas, as done in the above subsections for the log-gas on a contour.

Introduce $V(z) = {N \over |\Omega|} \int_\Omega \log | z - w| \, d^2w$ for the potential energy of the interaction between a unit charge and the background charge density supported on $\Omega$. Defining
$$
E_\Omega = - {1 \over 2 |\Omega|^2}
\int_\Omega d^2 z \int_\Omega d^2 w \,
\log | z - w| + {1 \over |\Omega|}
\int_\Omega \log |w| \, d^2 w,
$$
the construction of \S \ref{S1.1} for the Boltzmann factor of a one-component Coulomb system gives for the Boltzmann factor
\begin{equation}\label{U20}
e^{-\beta N^2 E_\Omega}
e^{-\beta \sum_{l=1}^N (V(z_l) - V(0))} \prod_{1 \le j < k \le N}
|z_k - z_j|^\beta.
\end{equation}
Since the free energy of a one-component Coulomb system constructed as in \S \ref{S1.1} is extensive in the thermodynamic limit, integrating over each $z_l \in \mathbb C^2$, dividing by $N!$, and taking the logarithm gives a quantity which is of order $N$. Now changing variables $z_l \mapsto \sqrt{N} Z_l$ tells us that for $N \to \infty$
\begin{align}
\begin{split}
\label{U21}
& \quad \log {1 \over N!} \int_{\mathbb C}
d^2 Z_1 \cdots \int_{\mathbb C}
d^2 Z_N \, e^{-\beta N \sum_{l=1}^N (
\tilde{V}(Z_l) - \tilde{V}(0))}
\prod_{1 \le j < k \le } | Z_k - Z_j|^\beta 
\\
&= - {\beta \over 4} N^2 \log N  + \beta N^2 E_\Omega +
\Big ( {\beta \over 4} - 1 \Big ) N \log N+ {\rm O}(N),
\end{split}
\end{align} 
where $\tilde{V}(Z) := \int_{\Omega_0} \log |Z - W| \, d^2 W$; cf.~(\ref{U11}). See also \cite{LS17}, \cite{AS21}, \cite{JV23}, and \cite{BSY25}, as well as the references therein, for further recent developments, and the earlier
works \cite{JMP94} ($\log N$ term) and
\cite{ZW06} (constant term).

As for checks and illustrations of (\ref{U21}), one first notes that the quantity $E_\Omega$ for $\Omega = B_R$,
$\Omega_0 = B_1$ is after evaluation, the constant term in (\ref{3.4b}) with $R$ eliminated according to $R=\sqrt{N/\pi}$. Hence, in the case $\beta = 2$, the expansion (\ref{U21}) is equivalent to that already checked in \S \ref{S3.4.2} for $\beta = 2$ in relation to GinUE. Similarly, for the choice of an elliptic domain
$\Omega_0 = \{(x_1,x_2): x_1^2/a_1^2 + x_2^2/a_2^2=1\}$, as it relates to $E_\Omega$ as specified by constant term in (\ref{3.4f}) with $a_1 \mapsto \sqrt{N} a_1, a_2 \mapsto \sqrt{N} a_2$. The corresponding leading asymptotics for $\beta = 2$ were verified in \S \ref{S3.4.2} for $\beta = 2$ in the context of  elliptic GinUE.

The (exterior) Green function formulas (\ref{U6}), (\ref{U7}), (\ref{U13}) all remain valid in the present setting of the 2d Coulomb gas with a background. Since the work \cite{Ja95}, this exterior Green function is known to be directly related to the surface component of the limiting (smoothed) two-point correlation function, $\sigma_\infty^T(\eta_1,\eta_2)$. In relation to the latter, introduce curvilinear coordinates $(\xi,\eta)$ in the directions normal and tangential to the surface of $\Omega$ at $(x,y)$. Let the Jacobian for this transformation be denoted $h(\xi,\eta)$. Then the correlation $\sigma_\infty^T(\eta_1,\eta_2)$ is specified by 
\begin{equation}\label{U22}
\sigma_\infty^T(\eta_1,\eta_2) = \lim_{N \to \infty} \int d \xi_1 \, h(\xi_1,\eta_1) 
\int d \xi_2 \, h(\xi_2,\eta_2) \rho_N^T((\xi_1,\eta_1),(\xi_2,\eta_2)).
\end{equation}
Beginning with a linear response type relation (\ref{U17}) (generalised from $B_1$ to $\Omega$, which requires replacing the final factor $\log |zw|$ by $\log |\zeta(z) \zeta(w)|$), it was deduced in \cite{Ja95} (see also \cite{CPRV89}) that
\begin{equation}\label{U23}
\sigma_\infty^T(\eta_1,\eta_2) =
- {1 \over \beta (2 \pi)^2 } 
{\partial^2 G((\xi_1,\eta_1),(\xi_2,\eta_2)) \over \partial \xi_1 \partial \xi_2},
\end{equation}
where the derivatives are evaluated at the surface.
The simplest example is the case of the unit disk, when $(\xi,\eta) = (r,\theta) $.
Substituting (\ref{U6}) for the Green function gives
\begin{equation}\label{U24}
\sigma_\infty^T(\theta_1,\theta_2) =
- {1 \over \beta 2  \pi^2 (2 R \sin (\theta_1 - \theta_2))/2)^2}
\end{equation}
(here we have written the final expression for the case that the domain $B_1$ is generalised to $B_R$). 

For $\beta = 2$, a direct verification of (\ref{U24}) from the definition (\ref{U22}) has been given in \cite{CPR87}. Also, notice that for $R=1$, $\beta =2$ this is the same functional form as that implied by (\ref{U16}), and thus $1/2$ of that for the smoothed two-point correlation for Haar distributed unitary random matrices. The slow, one on distance squared, decay along the boundary contrasts with the expected fast decay exhibited by the truncational two-point correlation in the bulk. For example, with $\beta =2$, the form of the latter in the thermodynamic limit is equal to $-\rho_{\rm b}^2 e^{-\pi \rho_{\rm b} |z_1 - z_2|^2}$
\cite{AJ81}.

The analogue of (\ref{U24}) in the case of an ellipse is \cite{FJ96}
\begin{equation}\label{U25}
\sigma_\infty^T(\eta_1,\eta_2) =
- {1 \over \beta 2  \pi^2 | e^{i \eta_1} - e^{i \eta_2}|^2 h(\xi_1,\eta_1) h(\xi_2,\eta_2)},
\end{equation}
with $\xi_1,\xi_2$ specified at the boundary. This can be shown to be consistent with (\ref{U22}), via an asymptotic analysis of
$\rho_N^T((\xi_1,\eta_1),(\xi_2,\eta_2))$ in the neighbourhood of the 
boundary \cite{FJ96}; see also the recent work \cite[Eq.~(67)]{O+24} where the latter is derived in the context of quantum Hall droplets.
A point of interest (not yet resolved) is to relate the functional form (\ref{U25}) to
the candidate for $\sigma_\infty^T(\eta_1,\eta_2)$ as given by $-1/\beta$ times (\ref{U18}) from \cite{AC23}, or more generally use (\ref{U23}) with $G$ given by (\ref{U8}) to deduce (\ref{U18});
the recent work \cite[supplementary material Eq.~(62) and (63) asserts this]{M+25}). Its validity leads to the (surface) fluctuation formula
(\ref{U19}), now multiplied by $1/\beta$ rather than $2/\beta$ as in the original setting of (\ref{U19}); recall the discussion of the paragraph following this latter equation.

We remark that generally a fluctuation formula for a linear statistic in the present uniform background model consists of the sum of a bulk contribution, and a surface contribution \cite{Fo99}. The latter is the subject of our present discussion and the former is given by $(1/2\pi \beta) \int_\Omega  \nabla  f \cdot
 \nabla  \bar{g} \, dx dy$. There are now many works relating to this; see
 \cite{RV07}, \cite{AHM11}, \cite{AHM15}, \cite{LS18},
  \cite{BBNY19},  \cite{Se23}, \cite{AC24}, \cite{ADM24}, \cite{ACC25}, \cite{ACC25a}.

 \begin{remark} ${}$ \\
 1.~In the circumstance of a large but finite system in which the thermodynamic limit is being taken, the interpretation given to (\ref{U22}) in \cite{Ja95} is as the large separation asymptotics of the surface correlation function. For example, (\ref{U23}) is applicable to a circle of (large) radius $R$ provided $\theta_1 - \theta_2$
 is independent of $R,N$. Similarly, in \cite{Ja95} (\ref{U22}) was used to predict the large separation asymptotic decay of the surface correlation along a boundaries of infinite extent (half plane, wedge). In particular, for the half plane plasma occupying $y<0$, it was deduced that $ \sigma_\infty^T(x_1,x_2)$ has the leading large $|x_1 - x_2|$ slow decay $-1/(2 \beta \pi^2 |x_1 - x_2|^2)$  \cite{Ja82}.
 \\
 2.~Applications of (\ref{U22}) to compute the asymptotic forms of surface correlations for three-dimensional Coulomb systems in the thermodynamic limit have been given in \cite{CPRV89},
 \cite{Ja95}. For example, in the case that the plasma occupies the half plane $z<0$, making use of (\ref{U10}) it was deduced that
 the surface charge correlation on the boundary $z=0$,
 $\sigma_\infty^T((x_1,y_1),(x_2,y_2))$, has for large separation of the points the leading asymptotic form
 $ -
 1 / (8 \beta \pi^2 
 ((x_1-x_2)^2 + (y_1-y_2)^2)^{3/2})$,
in agreement with
a result first obtained in \cite{Ja82}.
 
 \end{remark}

 \subsubsection{A Class of Generalised Fisher-Hartwig Type Asymptotic Expansions}
 As noted, the Boltzmann factor of the log-gas specified above (\ref{U12}) is proportional to the joint eigenvalue PDF for matrices from $U(N)$ with Haar distribution. With this latter viewpoint, it has long been realised that consideration of the particular average
 $\langle \prod_{l=1}^N e^{k f (\theta_l)} \rangle_{U(N)}$ is closely related to ${\rm Var} ( \sum_{l=1}^N f (\theta_l) )$ and thus (\ref{U14}) as one has the large $N$ expansion
 \begin{equation}\label{W2}
 \Big \langle \prod_{l=1}^N e^{k f (\theta_l)} \Big \rangle_{U(N)} \sim
 \exp \bigg ( k N \int_0^{2 \pi} f(\theta) \, d \theta +
 {k^2 \over 2} {\rm Var} \Big ( \sum_{l=1}^N f (\theta_l)\Big )+ {\rm o}(1)
 \bigg ).
\end{equation} 
This result is best known with the average on the LHS written as a Toeplitz determinant, and the variance for large $N$ expressed as the sum $\sum_{n=-\infty}^\infty |n| |f_n|^2 $ as noted in the text above
(\ref{U18}). In this form it is referred to as the strong Szeg\H{o}
theorem; see the review \cite{DIK13}. On the other hand, for its validity $f$ must be sufficiently smooth for the sum giving the limiting variance to converge. An example when this is not the case is $f(\theta) = - \log |e^{i \phi} - e^{i \theta} |^2$, for which
$|f_n|^2 = {1 \over n^2}$, $n \ne 0$, so the sum diverges logarithmically.

For the latter choice of $f$ we have $\langle \prod_{l=1}^N e^{k f (\theta_l)} \rangle_{U(N)} = \langle \prod_{l=1}^N  |
e^{i \phi} - e^{i \theta_l} |^{2k} \rangle_{U(N)}$. According to the $\beta = 2$ log-gas interpretation, the product on the RHS is a term in the Boltzmann factor due to a fixed charge of strength $k$ on the circle at angle $\phi$. As a one-component log-gas system with domain a circle of radius $R$, the uniform neutralising background must now have charge density
$- {N + k \over 2 \pi R}$. The total potential energy now consists of  background-background, background-particle, background-fixed charge, particle-particle and particle-fixed charge interactions, and is calculated as
 \begin{equation}\label{W3}
 U = \Big ( {N \over 2} - {k^2 \over 2} \Big ) \log R -
 {k } \sum_{j=1}^N \log | e^{i \phi} - e^{i \theta_j} | -
 \sum_{1 \le j < k \le N} \log |e^{i \theta_k} - e^{i \theta_j}|.
 \end{equation}
 For general $\beta > 0$,  in the thermodynamic limit $N,R \to \infty$ with $N/R$ fixed, requiring that the free energy be extensive and moreover that the next order term be ${\rm O}(1)$ as in keeping with the circle having no boundary, we are led to predict
 \begin{equation}\label{W4} 
 \Big \langle \prod_{l=1}^N  |
e^{i \phi} - e^{i \theta_l} |^{\beta k} \Big \rangle_{\beta, N}
\mathop{\sim}\limits_{N \to \infty} N^{-\beta k^2/2}.
 \end{equation}
 In the case $\beta =2$, when the LHS can be expressed as a 
 Toeplitz determinant, this is in keeping with the Fisher-Hartwig asymptotic formula; see e.g.~\cite[\S 14.5.3]{Fo10}. Its validity is known for general $\beta > 0$ making use of the fact that the average on the LHS is a particular Morris integral from the theory of the Selberg integral (\cite[Eq.~(4.4)]{Fo10}), and thus which has an explicit evaluation as a product of gamma functions. The RHS of (\ref{W4}) is quantifying the large $N$ divergence of the variance of the linear statistic $f(\theta) = - \log |e^{i \phi} - e^{i \theta} |^2$. In addition, a central limit theorem can be established
 \cite{BF98}, \cite{BD25}, as well as an explicit formula in terms of the Barnes-$G$ function for the next leading term (provided $\beta$ is a rational number) \cite{Fo92}. 

 It is shown in \cite{FF04} how the log-gas picture can make use of
 knowledge of the asymptotic formula (\ref{W4}), supplemented by the 
 knowledge of the next correction term to predict the explicit form  of $\Big \langle \prod_{l=1}^N e^{h(\theta_l)} |\prod_{j=1}^r
e^{i \phi_j} - e^{i \theta_l} |^{\beta k} \Big \rangle_{\beta, N}$
of the large $N$ form up to and including the constant term, 
for $h(\theta)$ smooth. In the case $\beta = 2$ this again relates to the Fisher-Hartwig asymptotic formula. The asymptotics of the analogous average for the GinUE can also be predicted by such considerations \cite{WW19}. The very recent work \cite{BDHK25} has given a proof and generalisation to allow for a wider class of normal matrix models.

\section{Balayage Measure and the Hole Probability}\label{S5}

\alert{For a uniformly charged domain $\Omega$, a question of interest is to determine a surface charge giving rise to the same potential outside of $\Omega$, which is referred to as the balayage measure. This is simple to calculate for a ball due to the rotational symmetry. In two-dimensions various other approaches are possible --- we give the case of the ellipse as an example. Application is given to calculating the leading 
large deviation asymptotics of the hole probability for no charges in a subdomain $\Omega_0$. While electrostatics also determines the leading asymptotics of the hole probability for the Coulomb potential in say one less dimension (e.g.~log-gas in dimension one), this is no longer determined by the balayage measure. On this, our account is restricted to providing references.}

\subsection{Balayage Measure}
Let $\Omega$, $V(\vec{r})$ be as in (\ref{1.3}). Let $\mu_{\partial \Omega}(\vec{r})$ be a charge density supported on $\partial \Omega$ creating a potential $V_{\partial \Omega}(\vec{r})$. This charge density is said to correspond to a balayage measure if
$V_{\partial \Omega}(\vec{r}) = V(\vec{r})$ for all $\vec{r}$ outside of $\Omega$ (continuity of the potential then gives that this equality holds true on the boundary as well). As a simple example, take $\Omega = B_R \subset \mathbb R^d$, which is to be filled with a uniform charge density of total charge $Q$. Then for
$\vec{r} \in \mathbb R^d \backslash \bar{B}_R$ we know from (\ref{3.2}) that for $\vec{r}$ outside of the ball $V(\vec{r}) = Q \Phi_d(\vec{r})$, which in words is $Q$ times the Coulomb potential.
On the other hand, with $c_d$ specified as below (\ref{1.1}), we know from (\ref{4.1}) and (\ref{4.2}) that the surface charge density $\sigma(\vec{r}) = Q/(R^{d-1} c_d)$ also gives rise to this same potential outside of the ball and hence corresponds to the balayage measure.

Distinct from the ball geometry is an annulus, which has both an inner and outer boundary. Let us consider this geometry, specified by $cR < |\vec{r}| < R$, $0 < c < 1$, which we take as being in $\mathbb R^2$ for definiteness. The potential due to a uniform background charge density of total charge $Q$ in the regions outside of $\Omega$ can be deduced from (\ref{3.4ai}) (which as written applies only for $\vec{r}$ inside of $\Omega$),
\begin{equation}\label{5.1}
V(\vec{r}) = \begin{cases}
- Q \log |\vec{r} |, & |\vec{r}| > R, \\
Q \Big ( {1 \over 2} - \log R + {c^2 \over 1 - c^2} \log c \Big ),
& |\vec{r}| < c R.
\end{cases}
\end{equation}
The balayage measure now consists of a charge density distributed over both boundaries, which using an obvious notation we write as $\sigma_R(\vec{r})+\sigma_{cR}(\vec{r})$. Introducing parameters $\alpha, \beta$ by setting $\sigma_R(\vec{r}) = Q \alpha/(2 \pi R)$, $\sigma_R(\vec{r}) = Q \beta/(2 \pi c R)$ and calculating the corresponding potentials shows $\alpha + \beta = 1$, $-\beta \log c =
{1 \over 2} + {c^2 \over 1 - c^2} \log c$, thus fully determining the balayage measure in this case.
We also refer the reader to \cite[Appendix~A]{ACCL24} for the derivation of the balayage measure associated with a general rotationally symmetric weight in $d=2$.

A moments based method was introduced in \cite{AR17} to calculate the balayage measure for several uniformly charged regions
$\Omega$ in two-dimensions which do not have a rotational symmetry --- further explicit examples were isolated in \cite{Ch23}.
We illustrate the moment method in the case of an ellipse.

\begin{proposition} 
Let $\Omega$ be the interior of the ellipse specified by the $d=2$ case of (\ref{3.5}). With the boundary parameterised by setting $x_1 = a_1 \cos \theta$, $x_2 = a_2 \sin \theta$, the surface charge density specifying the balayage measure is
\begin{equation}\label{5.2}
\sigma(\theta) =Q
{a_1 a_2 \over 2 \pi} \Big ( 1 - 
{a_1^2 - a_2^2 \over a_1^2 + a_2^2} \cos 2 \theta \Big ).
\end{equation}
\end{proposition}

\begin{proof}
The potential outside of a uniformly charged ellipse was not specified in our working relating to (\ref{3.13c}). Using elliptic coordinates, the explicit functional form can be found in \cite{VY10}, supplementing that of \cite{La32}. However this is not required for the present purpose, which proceeds instead by the $|z| \to \infty$ moment expansion
\begin{equation}\label{5.2a}
V(z) = Q \log |z|
+ Q \, {\rm Re} \sum_{l=1}^N {m_l  \over z^l},
\qquad 
m_l = \int_\Omega w^l \, d^2w.
\end{equation}
Writing $w = r \cos \theta + i r \sin \theta$, and setting $\alpha = {1 \over 2} (a_1+a_2)$, $\beta = {1 \over 2} (a_1 - a_2)$, a straightforward calculation shows $m_l$ is equal to  $ (\alpha^2-\beta^2)(2\alpha \beta)^{ l/2 } \frac{2}{l+2} \binom{l}{l/2} $  for $l$ even, and is equal to zero for $l$ odd. See \cite[Appendix~A]{ABO25} or \cite[Remark~2.5]{By24} for a derivation of this \alert{result} using the Joukowsky map in Prop.~\ref{P4.2}. Another straightforward calculation shows that these same (complex) moments are reclaimed from the specified surface charge, this implying equality of the potentials outside of $\Omega$.
\end{proof}

\subsection{Hole Probability}
Our viewpoint of a one-component charged system for the most part (an exception is \S \ref{S3.4A}) is that the background charge density is prescribed. \alert{Then}, from a physical perspective, \alert{as remarked several times throughout this review}, the charge density of the mobile charges must be to leading order equal and opposite that of the background so that the net electric field is zero --- if not the system would not be in equilibrium.

A possible circumstance is that there is a conditioning so that a region, $\Omega_0$ say, inside the droplet is free of mobile charges. The ratio of partition functions $Z_{\Omega \backslash \Omega_0} /
Z_\Omega$ then has the interpretation of the probability --- known as the \alert{hole} probability --- that in the system without conditioning, there are no mobile charges in $\Omega_0$. We know that the leading large $N$ form of the denominator $Z_\Omega$ is given in terms of certain electrostatic energies determined from the potential due to the interaction of the background with a particle. To apply the same reasoning to $Z_{\Omega \backslash \Omega_0}$, we need to be able to quantify, to leading order, the new particle charge density resulting from the conditioning. 

From an electrostatics viewpoint, a fundamental feature of interactions via the $d$-dimensional Coulomb potential in a $d$-dimensional domain is that the displaced charge \alert{will} accumulate on boundary $\partial \Omega_0$ \alert{of the excluded region}.
 Moreover, the rule that determines its distribution must  be that the potential experienced at a point in $\Omega \backslash \Omega_0$ due to the positive charge density of the displaced mobile charges, and the negative charge density of the background in $\Omega_0$, must exactly cancel (again, if not, the charges in this region would not be in equilibrium). These facts together tell us that the charge density on the boundary must correspond to the balayage measure with $Q = \rho_{\rm b} |\Omega_0|$ (the latter being the expected number of charges in the region $\Omega_0$ without conditioning).

In the case $\beta = 2$, $d=2$ a proof of this anticipated fact for a large class of $\Omega$ is given in \cite{AR17}. The proof was generalised to all $\beta > 0$ in \cite{Ad18}, but with the restriction that the background charge density be radially symmetric and $\Omega$ be a disk.
These latter restrictions were subsequently removed in \cite{Ch23}.

Consider now the charge neutral, particle free, system consisting of the background charge density $-\rho_{\rm b}$ in $\Omega_0$, and the oppositely charged balayage measure of charge density $\rho_{\rm b} \mu(\vec{r})$. Let $-\rho_{\rm b} U(\vec{r})$ denote the potential due to the background.
Without the minus sign, and restricted to $\partial \Omega_0$, this is also the potential due to the balayage measure. We thus have that the electrostatic energy due to interaction between the background and itself, between the balayage measure and itself, and between the background and the balayage measure is equal to
$
{\rho_{\rm b}^2 \over 2} \int_{\Omega_0} U(\vec{r}) \, d \vec{r}$, 
${\rho_{\rm b}^2 \over 2} \int_{\partial \Omega_0} U(\vec{r}) \mu(\vec{r})\, |d \vec{r}|$, $- \rho_{\rm b}^2 \int_{\partial \Omega_0} U(\vec{r}) \mu(\vec{r})\, |d \vec{r}|$ respectively. Denoting  the total electrostatic energy of this system by $\rho_{\rm b}^2 E_{\Omega_0}$ we therefore have
\begin{equation}\label{5.2b}
E_{\Omega_0} =
{1 \over 2} \Big ( \int_{\Omega_0} U(\vec{r}) \, d \vec{r} - \int_{\partial \Omega_0} U(\vec{r}) \mu(\vec{r})\, |d \vec{r}| \Big ).
\end{equation}
(Note that to compute this it is only necessary to know $U(\vec{r})$ up to an additive constant, as such a term cancels out.)

\begin{proposition}
(\cite{AR17}, \cite{Ad18}, \cite{Ch23}) 
We have
\begin{equation}\label{5.2c}
\lim_{\rho_{\rm b} \to \infty}
{1 \over \rho_{\rm b}^2} \log 
{Z_{\Omega \backslash \Omega_0} \over 
Z_\Omega} = - \beta E_{\Omega_0}.
\end{equation}
\end{proposition}
The hypothesis of such a result was put forward by Dyson \cite{Dy62} in the study of the gap probability in the circular unitary ensemble of Haar distributed unitary random matrices. Its application in random matrix settings more generally has become known as the Coulomb gas method. However, for the case of the log-potential confined to one-dimensional domains, it is no longer true that the equilibrium measure after conditioning involves the balayage measure. Rather the mobile charges removed from the gap form a continuous charge density in the remainder of the droplet. While this is an extra complexity, it turns out that it can be solved in many settings of interest in random matrix theory when the spectrum has a one-dimensional support (in the bulk as done by Dyson \cite{Dy62}, at the hard edge \cite{CM94},  and at the soft edge \cite{BDG01}, \cite{DM06}, \cite{DM08}, \cite{KC10}, \cite{RKC12}, \cite{MS14}). Moreover, in such a one-dimensional setting it has been possible to solve the electrostatic problem under the further conditioning that the gap contain a prescribed fraction of the total spectrum \cite{Dy95}, \cite{FS95},
\cite{FW12}, \cite{MNSV11}; see too the review \cite{Fo14}. 
For the one-dimensional Riesz gas, the problem of determining the constrained equilibrium measure has been studied in \cite{KKKMMS21,KKKMMS22}. 
For the one-component plasma in two dimensions with a constant background, this latter setting is only tractable when $\Omega_0$ is a disk centred on the origin \cite{Fo93x,ATW14}; see also \cite{ASZ14} for a related problem where the constraint is imposed on the half-plane.

\begin{remark}
As noted in the final paragraph of \S \ref{S1.1}
electrostatic arguments have been used in \cite{JLM93} to deduce the large $R$ form of the probability
that for the infinite two-dimensional one-component plasma, background charge density $-{1 \over \pi}$, the number of
particles is equal to 
$\lfloor \alpha R^\gamma \rfloor)$,
for $\gamma > {1 \over 2}$.
(The three dimensional case was considered from an electrostatics viewpoint too in \cite{JLM93}.) As a concrete example of the results, denoting this probability as 
${\rm Pr} (\mathcal N_R = \lfloor \alpha R^\gamma \rfloor)$, we have from
\cite{JLM93} in the cases $\gamma > 2$,
\begin{equation}\label{5.2d}
{\rm Pr} (\mathcal N_R = \lfloor \alpha R^\gamma \rfloor) \sim
e^{-{\beta \over 4} (\gamma - 2)
\alpha^2 R^{2 \gamma} \log R (1 + {\rm o(1)}};
\end{equation}
for a refinement in the case $\beta = 2$ see \cite{FL22}, and see too the discussion in \cite{GN18}, \cite{NY24}.
\end{remark}

\begin{acknowledgement}
Sung-Soo Byun was supported by the National Research Foundation of Korea grants (RS-2023-00301976, RS-2025-00516909).
Funding support to Peter Forrester for this research was through the Australian Research Council Discovery Project grant DP250102552. We thank the organisers of the MATRIX program ``Log-gases in Caeli Australi'', held in Creswick, Victoria, Australia during the first half of August 2025 
for the stimulation of the event leading to the present collaboration.
\end{acknowledgement}

\end{document}